%% file: main.tex
\documentclass[11pt]{article}
\usepackage[a4paper,margin=27mm]{geometry}
\usepackage[T1]{fontenc}
\usepackage{lmodern,microtype,amsmath,amssymb,amsthm,mathtools,booktabs,array}
\usepackage{tikz}
\usetikzlibrary{arrows.meta,positioning,calc}
\usepackage{algorithm,algpseudocode}
\usepackage{float,placeins,needspace}

\definecolor{emvdInk}{HTML}{24384A}
\definecolor{emvdBlue}{HTML}{365F86}
\definecolor{emvdTeal}{HTML}{247B76}
\definecolor{emvdAmber}{HTML}{B66B28}
\definecolor{emvdGray}{HTML}{A5B1BB}
\definecolor{emvdLight}{HTML}{F3F6F8}
\tikzset{
  emvdfig/.style={x=1cm,y=1cm,font=\small,text=emvdInk,
    line cap=round,line join=round,>=Latex},
  emvdhead/.style={anchor=west,font=\small\bfseries,text=emvdInk,inner sep=0pt},
  emvdnote/.style={font=\small,align=center,text=emvdInk,inner sep=1pt},
  emvdarrow/.style={-{Latex[length=1.8mm,width=1.3mm]},draw=emvdBlue,line width=.65pt},
  emvdguide/.style={draw=emvdGray,line width=.6pt,densely dashed},
  emvdpoint/.style={circle,fill=emvdTeal,draw=white,line width=.35pt,inner sep=0pt,minimum size=4.6pt},
  emvdsample/.style={circle,fill=emvdBlue,draw=white,line width=.35pt,inner sep=0pt,minimum size=5.5pt},
  emvderror/.style={circle,draw=emvdAmber,fill=white,line width=.95pt,inner sep=0pt,minimum size=5.3pt},
  emvdresult/.style={draw=emvdTeal!60!white,fill=emvdTeal!7,rounded corners=1mm,
    minimum height=7mm,text width=14.4cm,align=center,inner sep=2pt,font=\small}
}
\usepackage[hidelinks,pdftitle={The Approximation Complexity of Euclidean Metric Repair},pdfauthor={Chenglin Fan}]{hyperref}
\usepackage{aliascnt}
\usepackage[capitalise,noabbrev]{cleveref}
\allowdisplaybreaks[2]
\newtheorem{theorem}{Theorem}
\newaliascnt{lemma}{theorem}
\newtheorem{lemma}[lemma]{Lemma}
\aliascntresetthe{lemma}
\crefname{lemma}{Lemma}{Lemmas}
\Crefname{lemma}{Lemma}{Lemmas}
\newcommand{\R}{\mathbb R}
\newcommand{\Q}{\mathbb Q}
\newcommand{\E}{\mathbb E}
\newcommand{\Prb}{\mathbb P}
\newcommand{\OPT}{\operatorname{OPT}}
\newcommand{\err}{\operatorname{err}}
\newcommand{\rank}{\operatorname{rank}}
\newcommand{\eps}{\varepsilon}
\newcommand{\ind}{\mathbf 1}
\newcommand{\poly}{\operatorname{poly}}
\newcommand{\bin}[1]{\binom{#1}{2}}

\title{The Approximation Complexity of Euclidean Metric Repair}
\author{Chenglin Fan}
\date{}
\begin{document}
\maketitle
\begin{abstract}
In Euclidean Metric Violation Distance (EMVD), we are given pairwise squared distance observations on $n$ points and seek a realization in $\R^d$ that minimizes the number of modified observations. We give a deterministic $(1+\eps)$-approximation in time $f(d,\eps)(n+L)^{O(1)}$, where $L$ is the input encoding length and $f$ is a computable function depending only on $d$ and $\eps$. The algorithm allows coincident points and configurations of any affine dimension at most $d$. In contrast, when the dimension $d$ is part of the input, no deterministic polynomial-time constant-factor approximation exists unless $\mathsf P=\mathsf{NP}$. Moreover, for any fixed $0<\alpha<1$, approximation within $2^{(\log n)^{1-\alpha}}$ in deterministic polynomial time is impossible unless NP admits deterministic quasipolynomial-time algorithms. Both hardness results hold even when the observed squared distances between all points other than one are jointly realizable in $\R^d$. For arbitrary dimensions, we give a randomized polynomial-time $O(d+d^2/\log(2n))$-approximation. On a family of instances, a matching lower bound holds for our reconstruction algorithm even with polynomially many independent runs.
\end{abstract}

\input{intro}

\input{section2}
\input{section3}

\input{section4}

\section*{Conclusion}
Our results establish a separation between fixed and unrestricted dimensions for EMVD. A natural question is whether EMVD admits a polynomial-time $O(d)$-approximation in arbitrary dimensions. The lower bound in \cref{thm:pivotlower} shows that independent repetition of our reconstruction algorithm alone is insufficient. 

\section*{AI statement}
ChatGPT (OpenAI) was used to assist in developing,
checking, and refining parts of the proofs, as well
as in revising the manuscript.

\bibliographystyle{plain}
\bibliography{reference}

\end{document}

%% file: intro.tex
\section{Introduction}\label{sec:intro}
Pairwise distances often provide the only available information about the positions of a set of points. When some measurements are erroneous, a natural objective is to find a configuration that preserves as many observations as possible. Unlike metric nearness~\cite{brickell}, which minimizes a norm of the changes, this objective counts the number of modified distances, regardless of their magnitude.

For general metrics, this problem is known as \emph{sparse metric repair} or \emph{Metric Violation Distance} (MVD)~\cite{gilbertjain,fanmvd}. Fan et al.~\cite{fanmvd} establish hardness and give an $O(k^{1/3})$-approximation, where $k$ is the minimum number of modifications. Fan et al.~\cite{fangmvd} extend the problem to weighted graphs. Cohen-Addad et al.~\cite{cohenmetric} give an $O(\log n)$-approximation for metrics and a constant-factor approximation for ultrametrics. More recently, Bandyapadhyay et al.~\cite{bandy2026} obtain a linear kernel, a single-exponential parameterized algorithm, and an $O(\log(k+1))$-approximation for MVD. Etgar et al.~\cite{etgar2026} study its complexity on restricted graph classes.

We study the Euclidean counterpart, introduced by Bentert et al.~\cite{bentert} as \emph{Euclidean Metric Violation Distance} (EMVD). Given observed squared distances between $n$ points, the goal is to find positions in $\R^d$ that minimize the number of observations that must be changed. Formally, let $V=[n]$ and let $A\in\Q_{\ge0}^{n\times n}$ be symmetric with zero diagonal. For a configuration $X=(x_i)_{i\in V}\in(\R^d)^n$, define
\begin{equation}\label{eq:objective}
 \err_A(X)=\sum_{ij\in\bin V}\ind[A_{ij}\ne\|x_i-x_j\|^2],\qquad
 \OPT_d(A)=\min_{X\in(\R^d)^n}\err_A(X).
\end{equation}
Here $ij$ denotes an unordered pair of distinct point indices; different indices may have coincident positions. Let $L$ be the encoding length of $A$. We work in the bit model, representing each real algebraic coordinate by a univariate integer polynomial (given by its coefficient list) and a rational isolating interval. We use natural logarithms throughout. For the EPTAS, it suffices to supply a dyadic accuracy $\eps=2^{-j}$ for a nonnegative integer $j$, since any desired accuracy can be rounded down to such a value within a factor of two.

Even without a dimension bound, a metric need not admit a
Euclidean realization, so approximation algorithms for general
metric repair do not directly apply to EMVD.
Bentert et al.~\cite{bentert} show that EMVD is strongly NP-hard
even for $d=1$ and give an exact algorithm running in time
$(dk)^{O(d+k)}+n^{O(1)}$ in the real RAM model, where $k$ is
the edit budget. Their factor-two approximation applies to the
related problem of minimizing the number of deleted points,
leaving the approximability of EMVD open.
Fomin et al.~\cite{fomin} give fixed-parameter algorithms for
Euclidean distance matrix completion under several structural
parameterizations. Unlike EMVD, the completion problem specifies
in advance which entries are missing.
In this paper, we study the approximation complexity of EMVD
as a function of the dimension, without parameterizing the
running time by the edit budget.

\subsection{Our results}
Our main result is a deterministic efficient polynomial-time approximation scheme (EPTAS) for EMVD.
\begin{theorem}[Deterministic EPTAS]\label{thm:schemes}
There is a deterministic algorithm that, given an EMVD instance in dimension $d\ge1$ and an accuracy parameter $0<\eps\le1$, outputs a configuration $X\in(\R^d)^n$ satisfying
\[
 \err_A(X)\le (1+\eps)\OPT_d(A).
\]
The running time is $f(d,\eps)(n+L)^{O(1)}$ in the bit model, where $f$ is a computable function depending only on $d$ and $\eps$, and the polynomial exponent is an absolute constant.
\end{theorem}
The scheme applies to all edit budgets and permits coincident points and configurations lying in proper affine subspaces.

The next result shows a different picture when the dimension is unrestricted.
\begin{theorem}[Hardness in unbounded dimensions]\label{thm:hard}
When the dimension $d$ is part of the input, EMVD admits no deterministic polynomial-time approximation within any constant factor unless $\mathsf P=\mathsf{NP}$. Furthermore, for every fixed $0<\alpha<1$, a deterministic polynomial-time approximation within $2^{(\log n)^{1-\alpha}}$ would imply that NP admits deterministic quasipolynomial-time algorithms. Both hardness results hold even when all observed squared distances between points other than a single designated point are jointly realizable in $\R^d$ and $\OPT_d(A)=O(n)$.
\end{theorem}
Our third result gives an approximation algorithm whose running time is polynomial in the dimension itself. Its guarantee adapts to the affine dimension of a comparison configuration, which need not equal the ambient dimension.
\begin{theorem}[Dimension-dependent polynomial-time approximation]\label{thm:dim}
There is a randomized algorithm running in $\poly(n,d,L)$ time such that, for any fixed comparison configuration $X^*\in(\R^d)^n$ of affine dimension $r$, with probability at least $2/3$ its output $X$ satisfies
\[
 \err_A(X)\le O\!\left(r+1+\frac{(r+1)^2}{\log(2n)}\right)\err_A(X^*).
\]
A single reconstruction with exact uniform pivot sampling has expected cost at most $\frac12(r+2)(r+3)\err_A(X^*)$. In particular, the algorithm gives an $O(d+d^2/\log(2n))$-approximation in polynomial time, and an $O(d)$-approximation when $d=O(\log n)$.
\end{theorem}
The dimension dependence is tight on a family of instances for our reconstruction algorithm. For every fixed $C>0$, there are instances with $n=\Theta_C(d^2/\log d)$ on which the best of at most $n^C$ independent runs has approximation ratio $\Omega_C(d^2/\log d)$ with probability tending to one (\cref{thm:pivotlower}).

\subsection{Technical overview}

\paragraph{Additive approximation.}
The squared distance matrix of a configuration of affine dimension $r$ has rank at most $r+2$. This relates EMVD to $\ell_0$-low-rank approximation, for which Cohen-Addad et al.~\cite{cohenaddad} give a PTAS for fixed rank. Since low rank alone does not characterize Euclidean distance matrices, we work directly with point coordinates. Following techniques for dense constraint satisfaction problems~\cite{arora,cohenaddad}, we place the points block by block, using sampled reference positions to estimate the local cost of each placement. The references from earlier blocks have known descriptions, whereas the optimal positions in later blocks are unknown. To avoid a parameter-dependent polynomial exponent, we combine deterministic sampling of explicitly given geometric predicates, related to the techniques of Matou\v{s}ek~\cite{matousek95}, with a sample family for affine equality tests that is independent of their coefficients. We express the unknown reference coordinates as real variables and compress the resulting cost predicates. This yields a deterministic additive approximation with cost at most $\OPT_d(A)+\delta n^2$ in time $f(d,\delta)(n+L)^{O(1)}$.

\paragraph{From additive to multiplicative approximation.}
The additive guarantee suffices when the optimum is large. For smaller optima, we identify an affine subspace of an optimal configuration containing most of its points, such that every proper affine subspace omits many points relative to the number outside it. The additive solution contains a corresponding subspace with accurate reference data, which can be found among a list of candidates of size depending only on $d$ and $\eps$. For each candidate, we estimate each point's projection $u$ and squared distance $R$ to the subspace by voting over its references. Two distinct pairs $(u,R)$ agree on their predicted squared distances $\|u-v\|^2+R$ only on a proper affine hyperplane of reference positions $v$, or nowhere. For the reliable candidate, voting recovers the exact pair for all but $O_{d,\eps}(k/n)$ points, where $k=\OPT_d(A)$. Charging extra disagreements only to these points controls the additional cost by $O(\eps k)$. The remaining task is to fit inner products between vectors of prescribed norms in the orthogonal complement. We solve this problem recursively; after the transition to inner product fitting, every recursive call decreases the dimension. Both the recursion depth and the number of candidates at each node depend only on the parameters.

\paragraph{Hardness in unbounded dimensions.}
We reduce from minimizing the number of unsatisfied rational linear equations~\cite{abss,amaldi}. For an equation $a_i^\top x=b_i$, the identity $\|(x,t)-(a_i,0)\|^2-\|(x,t)\|^2=\|a_i\|^2-2a_i^\top x$ allows us to encode the equation by two squared distance observations. The extra coordinate $t$ lets us fix the norm of the point encoding the assignment without excluding an optimal assignment. To account for repairs that change the reference distances, we replicate the reference points and analyze a uniformly chosen representative from each group. This gives an exact relation between the two optima, and a separate extraction argument preserves the approximation ratio in the bit model.

\paragraph{Approximation in arbitrary dimensions.}
We construct an orthogonal coordinate frame by choosing random pivots. Each update preserves every point's earlier projection and its squared distance to the previous subspace, even if the observation used for the update is inconsistent. Distances to already resolved points are therefore fixed, allowing us to bound the propagation of errors and obtain an expected $O(d^2)$ approximation. To improve the guarantee, we compare the actual process with one that selects only pivots unaffected by erroneous observations. A reconstruction of cost at most $O(d+d^2/\log(2n))\OPT_d(A)$ occurs with probability $\Omega(1/\sqrt n)$. Taking the best of $O(\sqrt n)$ independent runs gives the desired ratio with probability at least $2/3$. A construction with errors confined to a perfect matching yields the lower bound for this reconstruction algorithm.

\Cref{sec:schemes} establishes the deterministic EPTAS,
\cref{sec:hard} proves the hardness results, and
\cref{sec:pivots} presents the approximation algorithm
for arbitrary dimensions and the corresponding lower bound.

%% file: section2.tex
\section{A deterministic EPTAS}\label{sec:schemes}

\paragraph{Problem formulation.}
The recursion for EMVD also solves an auxiliary problem, \emph{inner product fitting with prescribed norms}, which we call the prescribed-norm problem. Let $V=[n]$ be the point indices and $p$ the dimension of the current instance (initially $p=d$). In EMVD the inputs are squared distances $A_{ij}$ for $i<j$. In the prescribed-norm problem, we are given a squared norm $s_i\ge0$ for each $i\in V$ and a target inner product $g_{ij}$ for each $i<j$. We seek vectors $x_i\in\R^p$ with $\|x_i\|^2=s_i$ that minimize the number of unsatisfied inner product constraints. We set $A_{ji}=A_{ij}$, $g_{ji}=g_{ij}$, $A_{ii}=0$, and $g_{ii}=s_i$.

To treat both problems uniformly, define the feasible domain $\mathcal D_i$ of point $i$ and the binary loss $\ell_{ij}$ of pair $\{i,j\}$ by
\[
\begin{array}{lll}
\text{EMVD:} & \mathcal D_i=\R^p, &
\ell_{ij}(x,y)=\ind[\|x-y\|^2\ne A_{ij}],\\[2pt]
\text{Inner product fitting:} & \mathcal D_i=\{x\in\R^p:\|x\|^2=s_i\}, &
\ell_{ij}(x,y)=\ind[\langle x,y\rangle\ne g_{ij}].
\end{array}
\]
We assume that every domain $\mathcal D_i$ is nonempty. In particular, when $p=0$ in the prescribed-norm problem, we require $s_i=0$ for every $i\in V$. For a feasible configuration $X=(x_i)_{i\in V}$, define its cost by
\[
 \err(X)=\sum_{1\le i<j\le n}\ell_{ij}(x_i,x_j).
\]
Unless stated otherwise, let
\[
 k=\min_{X\in\prod_{i\in V}\mathcal D_i}\err(X)
\]
denote the optimum cost of the current instance, and let $X^*$ denote an optimal configuration. An optimum exists because the set of feasible costs is nonempty and finite.

In the prescribed-norm problem, a point with $s_i=0$ must be the origin. We remove all such points after counting the fixed cost of the unordered pairs involving them. We use affine subspaces for EMVD and linear subspaces for the prescribed-norm problem. Recursive instances may contain real algebraic numbers, represented as in \cref{sec:prelim}.

\paragraph{High-level idea.}
We first compute an additive approximation $Y$ of cost at most $k+\delta n^2$, where $\delta$ depends only on $p$ and $\eps$. If $k\ge\delta n^2/\eps$, this is already a $(1+\eps)$-approximation. Otherwise, the squared distances or inner products induced by $Y$ and $X^*$ differ on at most $2k+\delta n^2$ pairs. From $Y$ we enumerate a collection of subspaces of size depending only on $p,\eps$, one of which provides reliable reference positions. Voting on this candidate recovers the projections and squared distances to the subspace for all but $O_{p,\eps}(k/n)$ points, and the selected pairs admit a completion of cost at most $(1+\eps/4)k$. The residual vectors have prescribed norms, so their inner products can be fitted recursively. We retain $Y$ along with every completed candidate and return the configuration of minimum cost; the algorithm need not know $k$.

\paragraph{Algorithm.}
Let $I$ denote the current instance in dimension $p$, and let $0<\eps\le1$ be the required accuracy. Define
\begin{equation}\label{eq:threshold}
 \begin{gathered}
 a=\eps/4096,\qquad \tau_{-1}=1,\qquad \tau_j=a^{j+1}\ (0\le j<p),\qquad \tau_p=0,\\
 \theta=\min_{0\le j\le p}(\tau_{j-1}-\tau_j),\qquad
 \gamma=\frac{\eps\theta^2}{8192(p+1)}.
 \end{gathered}
\end{equation}
Set
\begin{equation}\label{eq:eptas-parameters}
 \gamma'=\gamma/8,\qquad \delta=\eps\gamma',\qquad \eta=\eps\theta/512.
\end{equation}
The thresholds $\tau_j$ control concentration in subspaces, $\theta$ is the smallest gap between consecutive thresholds, and $\gamma'$ separates the two regimes in the analysis. The additive approximation and voting use accuracies $\delta$ and $\eta$, respectively.

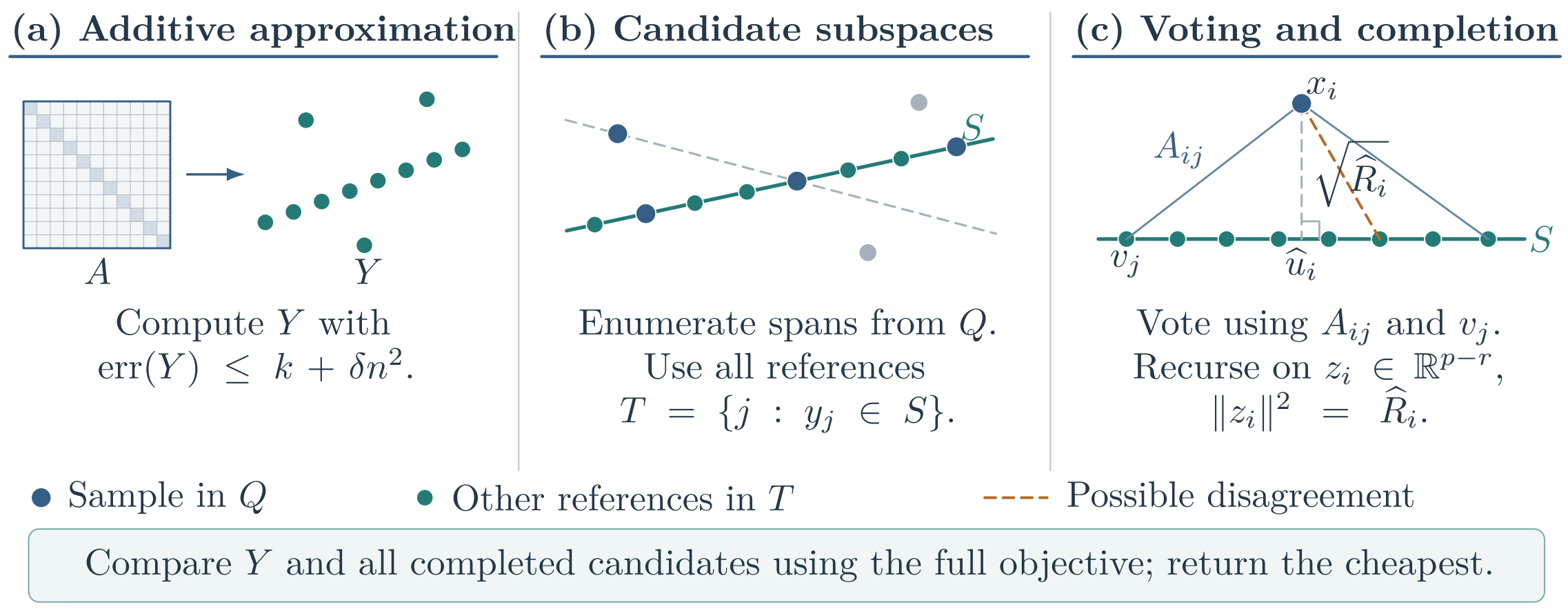
\begin{figure}[!htbp]
\centering
\begin{tikzpicture}[emvdfig]
\draw[emvdGray!65,line width=.45pt] (4.98,.18)--(4.98,-4.21);
\draw[emvdGray!65,line width=.45pt] (10.08,.18)--(10.08,-4.21);
\begin{scope}
  \node[emvdhead] at (.10,0) {(a) Additive approximation};
  \draw[emvdBlue,line width=.85pt] (.10,-.24)--(4.77,-.24);
  \begin{scope}[shift={(.23,-.67)},x=.128cm,y=.128cm]
    \fill[emvdLight] (0,0) rectangle (11,-11);
    \foreach \i in {0,...,10}{\fill[emvdBlue!22] (\i,-\i) rectangle ++(1,-1);}
    \draw[step=1,emvdGray!75,line width=.25pt] (0,-11) grid (11,0);
    \draw[emvdBlue,line width=.55pt] (0,0) rectangle (11,-11);
  \end{scope}
  \node[emvdnote] at (.94,-2.30) {$A$};
  \draw[emvdarrow] (1.80,-1.37)--(2.36,-1.37);
  \foreach \x in {2.55,2.82,3.09,3.36,3.63,3.90,4.17,4.44}
    \node[emvdpoint] at (\x,{-1.83+.37*(\x-2.55)}) {};
  \foreach \x/\y in {2.94/-.85,3.50/-2.05,4.10/-.65}
    \node[emvdpoint] at (\x,\y) {};
  \node[emvdnote] at (3.54,-2.30) {$Y$};
  \node[emvdnote,text width=4.6cm,anchor=north] at (2.43,-2.62)
    {Compute $Y$ with\\$\err(Y)\le k+\delta n^2$.};
\end{scope}
\begin{scope}[shift={(5.10,0)}]
  \node[emvdhead] at (.10,0) {(b) Candidate subspaces};
  \draw[emvdBlue,line width=.85pt] (.10,-.24)--(4.77,-.24);
  \pgfmathsetmacro{\altY}{-1.91+(.88/4.09)*(2.55-.34)}
  \pgfmathsetmacro{\altSlope}{(\altY+.98)/(2.55-.83)}
  \draw[emvdguide] (.33,{-.98+\altSlope*(.33-.83)})--(4.46,{-.98+\altSlope*(4.46-.83)});
  \draw[emvdTeal,line width=1pt] (.34,-1.91)--(4.43,-1.03);
  \foreach \x in {.61,1.10,1.57,2.07,2.55,3.04,3.55,4.08}
    \node[emvdpoint] at (\x,{-1.91+(.88/4.09)*(\x-.34)}) {};
  \foreach \x in {1.10,2.55,4.08}
    \node[emvdsample] at (\x,{-1.91+(.88/4.09)*(\x-.34)}) {};
  \node[emvdsample] at (.83,-.98) {};
  \node[circle,fill=emvdGray,inner sep=0pt,minimum size=4.6pt] at (3.23,-2.12) {};
  \node[circle,fill=emvdGray,inner sep=0pt,minimum size=4.6pt] at (3.72,-.68) {};
  \node[anchor=south west,font=\small,text=emvdTeal,inner sep=1pt] at (4.07,-1.08) {$S$};
  \node[emvdnote,text width=4.6cm,anchor=north] at (2.43,-2.62)
    {Enumerate spans from $Q$.\\Use all references\\$T=\{j:y_j\in S\}$.};
\end{scope}
\begin{scope}[shift={(10.20,0)}]
  \node[emvdhead] at (.10,0) {(c) Voting and completion};
  \draw[emvdBlue,line width=.85pt] (.10,-.24)--(4.77,-.24);
  \draw[emvdTeal,line width=1pt] (.34,-1.99)--(4.43,-1.99);
  \foreach \x in {.61,1.10,1.57,2.07,2.55,3.04,3.55,4.08}
    \node[emvdpoint] at (\x,-1.99) {};
  \draw[emvdBlue!75,line width=.55pt] (.61,-1.99)--(2.29,-.69)--(4.08,-1.99);
  \draw[emvdAmber,densely dashed,line width=.75pt] (2.29,-.69)--(3.04,-1.99);
  \draw[emvdguide] (2.29,-.69)--(2.29,-1.99);
  \draw[emvdGray,line width=.55pt] (2.29,-1.82)--(2.46,-1.82)--(2.46,-1.99);
  \node[emvdsample] at (2.29,-.69) {};
  \node[font=\small,anchor=south west,inner sep=1pt] at (2.29,-.69) {$x_i$};
  \node[font=\small,anchor=west,inner sep=1pt] at (2.35,-1.36) {$\sqrt{\widehat R_i}$};
  \node[emvdnote] at (2.29,-2.23) {$\widehat u_i$};
  \node[emvdnote,text=emvdBlue] at (1.12,-1.14) {$A_{ij}$};
  \node[emvdnote] at (.61,-2.23) {$v_j$};
  \node[anchor=west,font=\small,text=emvdTeal,inner sep=1pt] at (4.43,-1.99) {$S$};
  \node[emvdnote,text width=4.6cm,anchor=north] at (2.43,-2.62)
    {Vote using $A_{ij}$ and $v_j$.\\Recurse on $z_i\in\R^{p-r}$,\\$\|z_i\|^2=\widehat R_i$.};
\end{scope}
\node[emvdsample] at (.40,-4.47) {};
\node[anchor=west,font=\small] at (.52,-4.47) {Sample in $Q$};
\node[emvdpoint] at (4.08,-4.47) {};
\node[anchor=west,font=\small] at (4.20,-4.47) {Other references in $T$};
\draw[emvdAmber,densely dashed,line width=.75pt] (9.45,-4.47)--(10.05,-4.47);
\node[anchor=west,font=\small] at (10.10,-4.47) {Possible disagreement};
\node[emvdresult] at (7.55,-5.12)
  {Compare $Y$ and all completed candidates using the full objective; return the cheapest.};
\end{tikzpicture}
\caption{The EPTAS for EMVD. (a) Compute an additive solution $Y$. (b) Enumerate subspaces spanned by sampled points (blue). For each candidate $S$, all points of $Y$ in $S$ serve as references $T$. (c) Vote using the original observations and reference positions, then optimize the residual vectors recursively. The amber segment illustrates a possible disagreement.}
\label{fig:eptas-roadmap}
\end{figure}

\Cref{alg:eptas,fig:eptas-roadmap} summarize the algorithm. After removing zero norm points in the prescribed-norm problem, we solve the base cases exactly (\cref{sec:prelim}) and otherwise compute $Y=(y_i)_{i\in V}$ using \cref{lem:detadd}.

To generate candidate subspaces, we apply \cref{lem:detcompress} separately to the two families of predicates
\[
 \ind[y_i\in S],\qquad \ind[y_i\in S\setminus S'],\qquad i\in V,
\]
where $S,S'$ range over affine or linear subspaces, according to the problem type. Both applications use accuracy $\theta/4$. Let $Q\subseteq V$ be the union of the sampled index sets. For EMVD we enumerate the affine spans of nonempty tuples of at most $p+1$ sampled points. For the prescribed-norm problem we enumerate the positive-dimensional linear spans of at most $p$ sampled points.

For each candidate subspace $S$, let $r=\dim S$ and choose an origin $o_S\in S$, taking $o_S=0$ in the prescribed-norm problem. Using Gram--Schmidt orthogonalization, construct matrices $U_S\in\R^{p\times r}$ and $W_S\in\R^{p\times(p-r)}$ whose columns form orthonormal bases of the direction space of $S$ and its orthogonal complement, respectively. Let
\[
 T=\{j\in V:y_j\in S\}
\]
be the reference set, and let $v_j=U_S^\top(y_j-o_S)\in\R^r$ denote the coordinates of reference point $j$ for $j\in T$.

For each point $i$, we seek projection coordinates $u\in\R^r$ and a squared distance $R$ to $S$. A pair $(u,R)$ is feasible if $R\ge0$, with $R=0$ whenever $r=p$. In the prescribed-norm problem, we additionally require $\|u\|^2+R=s_i$. Using the observations of the current instance, define the voting cost
\begin{equation}\label{eq:voting-cost}
 c_i(u,R)=\begin{cases}
 \displaystyle\sum_{j\in T\setminus\{i\}}\ind[\|u-v_j\|^2+R\ne A_{ij}],&\text{EMVD},\\[4pt]
 \displaystyle\sum_{j\in T\setminus\{i\}}\ind[\langle u,v_j\rangle\ne g_{ij}],&\text{inner product fitting}.
 \end{cases}
\end{equation}
For each $i\in V$, we find a feasible pair $(\widehat u_i,\widehat R_i)$ satisfying
\begin{equation}\label{eq:approxvote}
 c_i(\widehat u_i,\widehat R_i)\le\min_{u,R}c_i(u,R)+\eta n,
\end{equation}
where the minimum is over feasible pairs. The computation is described in \cref{sec:concentration}.

Once the projections and squared distances are fixed, the remaining problem is a prescribed-norm instance $I_S$ in dimension $p-r$, with squared norms $\widehat R_i$ and target inner products
\begin{equation}\label{eq:residual}
 \widetilde g_{ij}=\begin{cases}
 \displaystyle\frac{\|\widehat u_i-\widehat u_j\|^2+\widehat R_i+\widehat R_j-A_{ij}}{2},&\text{EMVD},\\[6pt]
 g_{ij}-\langle\widehat u_i,\widehat u_j\rangle,&\text{inner product fitting}.
 \end{cases}
\end{equation}
If $r<p$, we solve $I_S$ recursively with accuracy $\eps/4$ to obtain vectors $z_i\in\R^{p-r}$ satisfying $\|z_i\|^2=\widehat R_i$. If $r=p$, feasibility implies $\widehat R_i=0$ for every $i$, and we set all residual vectors to zero without a recursive call. The completed configuration $X_S=(x_i)_{i\in V}$ is given by
\begin{equation}\label{eq:ambient-reconstruction}
 x_i=o_S+U_S\widehat u_i+W_Sz_i.
\end{equation}
We evaluate the cost of every completed configuration on all unordered pairs and return the configuration of minimum cost among $Y$ and all $X_S$, restoring any previously removed points at the origin.

\begin{algorithm}[!ht]
\caption{Deterministic EPTAS for EMVD and inner product fitting}
\label{alg:eptas}
\small
\begin{algorithmic}[1]
\Require Instance $I$ in dimension $p$, accuracy $0<\eps\le1$
\Ensure Feasible configuration of cost at most $(1+\eps)k$
\Function{Solve}{$I,p,\eps$}
  \State Set $\theta,\delta,\eta$ using \eqref{eq:threshold} and \eqref{eq:eptas-parameters}.
  \State Remove zero norm points for inner product fitting; update $I,V,n$.
  \If{$n\le1$ or $n<4/\theta$ or $p=0$}
    \State Solve exactly (\cref{sec:prelim}); restore removed points and return.
  \EndIf
  \State $Y\gets\Call{AdditiveApproximation}{I,p,\delta}$ \Comment{\cref{lem:detadd}}
  \State $X_{\mathrm{best}}\gets Y$
  \State Compress the predicates $\ind[y_i\in S]$ and $\ind[y_i\in S\setminus S']$ separately
  \Statex \hspace{3em} at accuracy $\theta/4$; let $Q$ be the union of their supports.
  \State Enumerate spans $\mathcal S$ of nonempty tuples indexed by $Q$:
  \Statex \hspace{3em} affine spans of at most $p+1$ points for EMVD;
  \Statex \hspace{3em} positive-dimensional linear spans of at most $p$ points otherwise.
  \ForAll{$S\in\mathcal S$}
    \State $r\gets\dim S$; construct the frame $(o_S,U_S,W_S)$.
    \State $T\gets\{j\in V:y_j\in S\}$; set $v_j\gets U_S^\top(y_j-o_S)$ for $j\in T$.
    \State For each $i\in V$, find feasible $(\widehat u_i,\widehat R_i)$ using \eqref{eq:approxvote}.
    \If{$r=p$}
      \State Set $z_i=0$ for all $i$.
    \Else
      \State Form $I_S$ using \eqref{eq:residual} and squared norms $\widehat R_i$.
      \State $(z_i)_{i\in V}\gets\Call{Solve}{I_S,p-r,\eps/4}$
    \EndIf
    \State Construct $X_S$ using \eqref{eq:ambient-reconstruction}.
    \If{$\err(X_S)<\err(X_{\mathrm{best}})$} \Comment{Score all pairs}
      \State $X_{\mathrm{best}}\gets X_S$
    \EndIf
  \EndFor
  \State Restore removed points at the origin in $X_{\mathrm{best}}$.
  \State \Return $X_{\mathrm{best}}$
\EndFunction
\end{algorithmic}
\end{algorithm}

Every recursive call for the prescribed-norm problem strictly decreases the dimension. EMVD may first produce a prescribed-norm instance of the same dimension. We establish the algebraic and sampling tools in \cref{sec:prelim,sec:detcompress}, the additive approximation in \cref{sec:detadd}, and the voting guarantee in \cref{sec:concentration}. The proof of the EPTAS is in \cref{sec:detmult}.

\subsection{Exact algebraic computation and local optimization}\label{sec:prelim}

\paragraph{Exact Euclidean realizability.}
Given a complete squared distance matrix $A$, choose an anchor point $a\in V$ and define the $(n-1)\times(n-1)$ Gram matrix $K$ by
\begin{equation}\label{eq:gram}
 K_{ij}=\frac12(A_{ai}+A_{aj}-A_{ij}),\qquad i,j\ne a.
\end{equation}
For any realization of $A$, we have $K_{ij}=\langle x_i-x_a,x_j-x_a\rangle$. Thus, $A$ is realizable in $\R^p$ if and only if $K\succeq0$ and $\rank K\le p$. Conversely, a factorization $K=XX^\top$ gives a realization with $x_a=0$. For rational input, positive semidefiniteness, rank, and an $LDL^\top$ factorization of $K$ can be computed in polynomial time in the input encoding length.

\paragraph{Real algebraic computation.}
We use quantifier elimination and sign-condition sampling for polynomial systems~\cite{basu}. Consider a formula over the reals with at most $q$ variables, size at most $M$, and polynomials of degree at most $D$ with integer coefficients of bit length at most $B$. Rational coefficients can be converted to integers by clearing positive denominators. The computation takes time
\[
 f(q,D,M)(B+1)^c,
\]
where $f$ is a computable function and $c$ is an absolute constant. The resulting polynomial descriptions and algebraic sample points have degrees bounded in terms of $q,D,M$ and coefficient bit lengths at most $O_{q,D,M}(B+1)$. The isolating intervals can also be represented with bit lengths bounded in this way.

For algebraic coefficients, introduce a variable for each coefficient, together with its defining polynomial and rational isolating interval, and eliminate these variables. If their number and individual degrees are bounded, the increase in formula complexity depends only on the parameters. Thus local computations do not require a common number field containing the coordinates of all $n$ points.

\paragraph{Exact optimization of small instances.}
If $p=0$, all feasible points are at the origin, so their cost is fixed. If $n$ is bounded in terms of the parameters, enumerate subsets of pairwise constraints in decreasing cardinality and test their joint feasibility with $x_i\in\mathcal D_i$. A configuration satisfying the first feasible subset is optimal. The number of subsets, variables, and constraints depends only on the parameters, so the algebraic tools give an exact algorithm with an absolute polynomial exponent.

\paragraph{VC dimension of geometric predicates.}
Uniform convergence for range spaces of bounded VC dimension~\cite{vc} allows us to approximate the average of a collection of predicates by a small sample. Boolean combinations of a bounded number of polynomial sign tests, with bounded degree and a bounded number of real parameters, define classes of bounded VC dimension~\cite{goldberg}. This includes equality tests. Unless stated otherwise, these complexity bounds depend only on the dimension and accuracy parameters and, for algebraic data, the bound on individual degrees.

\paragraph{Local optimization.}
Given a point $i$, a multiset $J$ of $M$ reference indices, fixed reference positions $v_j\in\R^p$, and nonnegative weights $w_j$ for the reference occurrences, consider the local optimization problem
\[
 \min_{x\in\mathcal D_i}\sum_{j\in J}w_j\ell_{ij}(x,v_j).
\]
We call a minimizer an optimal local placement.

\begin{lemma}[Local representatives]\label{lem:local}
For either geometric problem, an optimal local placement can be found among $O((M+1)^{p+1})$ candidates. Each candidate is chosen by a fixed canonical rule from the feasible set of the domain constraint and at most $p+1$ reference constraints. If $M$ is bounded in terms of the parameters, the optimal local placement, with a fixed rule for breaking ties, has a semialgebraic description whose complexity depends only on the parameters.
\end{lemma}
\begin{proof}
Let $x^*\in\mathcal D_i$ be an optimal local placement and consider the reference constraints satisfied by $x^*$. For EMVD, retain one satisfied constraint of the form $\|x-v_j\|^2=A_{ij}$ and subtract it from every other satisfied constraint. The resulting affine linear system is consistent and has a basis of at most $p$ equations. The retained constraint, together with the original constraints corresponding to this basis, defines the same feasible set as all the reference constraints satisfied by $x^*$. For the prescribed-norm problem, the reference constraints are already linear, so at most $p$ of them suffice in addition to the prescribed norm constraint. If $x^*$ satisfies no reference constraint, we use the empty subset.

We enumerate all subsets of at most $p+1$ reference constraints and discard those that are infeasible together with the domain constraint. For each remaining subset, choose a canonical point by first minimizing $\|x\|^2$ over its feasible set and then lexicographically minimizing the coordinates among the norm minimizers. Such a point exists because the feasible set is closed and the squared norm has compact sublevel sets. At least one of the resulting candidates satisfies every reference constraint satisfied by $x^*$. Since the weights are nonnegative, this candidate also attains the optimal local cost.

The number of subsets is $O((M+1)^{p+1})$. When $M$ is bounded, the feasibility tests, the canonical selection rule, and the comparison of weighted costs can all be expressed by first-order formulas over the reals of bounded size and degree, with the reference coordinates, observations, and weights as parameters. Quantifier elimination gives the required semialgebraic description. Ties between candidates are broken according to a fixed enumeration order.
\end{proof}

\subsection{Deterministic sampling}\label{sec:detcompress}

The additive scheme uses two sampling tools. The first compresses explicitly given predicates, as needed for points already placed. The second constructs a family of samples for predicates with unknown coefficients, as needed for points whose optimal positions are not known.

\begin{lemma}[Deterministic predicate compression]\label{lem:detcompress}
Let $h_1,\ldots,h_N:\R^a\to\{0,1\}$ be binary predicates, each described by a Boolean formula with at most $s$ polynomial sign tests of degree at most $D$. Suppose that the coefficients are rational and have bit length at most $B$. For $N\ge1$ and any $0<\eta\le1$, one can deterministically find a multiset $J$ of indices whose size depends only on $a,s,D,\eta$ and satisfies
\begin{equation}\label{eq:detcompress}
 \sup_{z\in\R^a}\left|\frac1N\sum_{i=1}^N h_i(z)-\frac1{|J|}\sum_{i\in J}h_i(z)\right|\le\eta.
\end{equation}
The running time is $f(a,s,D,\eta)(N+B+1)^c$, where $c$ is an absolute constant. The same statement holds for algebraic coefficients of bounded individual degrees and encoding lengths at most $B$, with the degree bound included among the parameters.
\end{lemma}

\begin{proof}
\textbf{Enumerating realizable patterns.}\enspace
For any collection of $m$ predicates, consider the binary vectors
\[
 (h_1(z),\ldots,h_m(z)),\qquad z\in\R^a.
\]
The number of distinct vectors is at most $(c_0(m+1))^{c_0}$, where $c_0$ depends only on $a,s,D$. Moreover, these vectors can be enumerated by computing sample points in the realizable sign conditions of the underlying $ms$ polynomials. By the standard bounds for real algebraic computation~\cite{basu}, this takes $(m+1)^{c_1}(B+1)^c$ time, where $c_1$ depends only on $a,s,D$ and $c$ is an absolute constant, up to factors depending only on the parameters. The intermediate polynomials have degrees bounded in terms of the parameters and coefficient bit lengths at most $O_{a,s,D}(B+1)$. For algebraic coefficients, we first eliminate their defining variables separately for each predicate, as described in \cref{sec:prelim}. This increases the formula complexity only by an amount depending on the parameters and avoids constructing a common number field.

\par\smallskip\noindent\textbf{Recursive compression.}\enspace
The case $N=1$ is immediate. For $N\ge2$, arrange the $N$ indices as the leaves of a balanced binary tree of height $H=\lceil\log_2 N\rceil$. Set $\xi=\eta/(2H)$ and
\[
 T=\left\lceil C_0\eta^{-2}(1+\lceil\log_2(N+1)\rceil)^4\right\rceil,
\]
where $C_0$ is a sufficiently large positive integer depending only on $a,s,D$. At each node, we represent the uniform distribution over its descendant leaves by a small multiset of indices. A leaf is represented by its own index. At an internal node, take the mixture of the two child distributions, weighted by their respective numbers of descendant leaves. The mixture has support of size at most $2T$. We compress it into a multiset of exactly $T$ indices.

Enumerate all realizable binary patterns on the support of the mixture. For any such pattern, let $\mu$ be its mean under the mixture. If we draw $T$ independent indices from the mixture, Hoeffding's inequality gives
\[
 \Prb\left[\left|\frac1T\sum_{t=1}^T X_t-\mu\right|>\xi\right]
 \le 2\exp(-2T\xi^2),
\]
where $X_t$ is the binary value of the pattern at the $t$-th sampled index. Since there are at most $(c_0(2T+1))^{c_0}$ realizable patterns, the choice of $C_0$ ensures that the expected number of patterns violating the error bound is less than one.

We construct the sample deterministically using the method of conditional expectations. After fixing a prefix of the sample, the conditional probability that a particular pattern violates the error bound is given by a binomial tail probability for the remaining draws, with the number of ones already observed included in the threshold. This probability can be computed exactly using rational arithmetic. We sum these probabilities over all realizable patterns and choose the next index so that the sum does not increase. Such a choice always exists because the current sum is the weighted average of the conditional sums over all possible next indices. After all $T$ indices have been selected, the sum equals the number of violating patterns. Since it is an integer smaller than one, no pattern violates the error bound.

The mixture probabilities have a common denominator at most $NT$. The binomial probabilities and their sums can therefore be computed using numbers of bit length polynomial in $T$ and $\log(N+1)$. Together with the pattern enumeration, each merge takes
\[
 (1+\log N)^{f(a,s,D,\eta)}(B+1)^c
\]
time, up to factors depending only on the parameters. The error of the mixture is at most the larger of the errors of its two child distributions, and the compression introduces an additional error of at most $\xi$. By induction on the height of the tree, the multiset at the root approximates the original uniform distribution within $H\xi=\eta/2$, simultaneously for all $z\in\R^a$.

\par\smallskip\noindent\textbf{Final compression.}\enspace
The root multiset has size $T$, which still depends on $N$. To remove this dependence, consider the family of binary functions on the indices defined by $i\mapsto h_i(z)$ for $z\in\R^a$. Its VC dimension is bounded in terms of $a,s,D$~\cite{goldberg}. Uniform convergence therefore guarantees a multiset of $v=v(a,s,D,\eta)$ draws from the root multiset that approximates its average within $\eta/2$, uniformly over all $z$.

We enumerate all $T^v$ ordered samples of length $v$ from the root multiset and test each sample against all realizable patterns on the root support. This finds a sample with the required error in time
\[
 (1+\log N)^{f(a,s,D,\eta)}(B+1)^c,
\]
up to factors depending only on the parameters. The triangle inequality gives the error bound in \eqref{eq:detcompress}.

Finally, the binary tree has $O(N)$ nodes, and for every fixed $u>0$ there is a constant $C_u$ such that
\[
 (1+\log N)^u\le C_uN
\]
for all $N\ge1$. Hence the total running time has an absolute polynomial exponent in $N+B+1$. The algorithm returns only a multiset of indices and introduces no new algebraic coefficients into the sampled predicates.
\end{proof}

For predicates whose coefficients are unknown, the following sample family gives a uniform error guarantee in expectation using pairwise independent sampling.

\begin{lemma}[A universal sample family for affine equalities]\label{lem:universal}
For every $a$ and $0<\eta\le1$, there is an integer $v=v(a,\eta)$ with the following property. Let $N\ge v$, and let $P$ be a prime satisfying $N\le P<2N$. Identify the $N$ indices with distinct elements of $\mathbb F_P$ and treat the remaining elements as dummy indices.

There is an explicitly enumerable family $\mathcal F$ of $P^2$ ordered samples of length $v$ such that, for every collection of affine functions $\ell_j:\R^a\to\R$, indexed by $j\in\mathbb F_P$, the predicates
\[
 h_j(z)=\ind[\ell_j(z)\ne0]
\]
satisfy
\[
 \E_{J\sim\mathcal F}\left[\sup_{z\in\R^a}
 \left|\frac1v\sum_{j\in J}h_j(z)-\frac1P\sum_{j\in\mathbb F_P}h_j(z)\right|\right]\le\eta,
\]
where $J$ is chosen uniformly from $\mathcal F$ and the sample sum counts occurrences with multiplicity. The same guarantee holds if a fixed subset of indices, including the dummy indices, is assigned the constant zero predicate. The family $\mathcal F$ does not depend on the affine functions.
\end{lemma}

\begin{proof}
\textbf{A finite collection of brackets.}\enspace
For $z\in\R^a$, write
\[
 h(z)=(h_j(z))_{j\in\mathbb F_P}\in\{0,1\}^P.
\]
A bracket $[l,u]$, where $l,u\in\{0,1\}^P$ and $l\le u$ coordinatewise, consists of all binary vectors lying coordinatewise between $l$ and $u$. Its width is
\[
 \frac1P\sum_{j\in\mathbb F_P}(u_j-l_j).
\]
We show that, for any $0<\beta\le1$, all realizable vectors $h(z)$ can be covered by at most $C(a,\beta)$ brackets of width at most $\beta$.

For a simplex $\sigma$ of dimension at most $a$, define its crossing set to consist of the indices $j$ for which $\ell_j$ takes both strictly positive and strictly negative values at the vertices of $\sigma$. The family of crossing sets has bounded VC dimension: a simplex is described by at most $a+1$ vertices, and membership in its crossing set is determined by a Boolean combination of a bounded number of linear sign tests. By the VC $\eps$-net theorem, there is a set of $f(a,\beta)$ indices intersecting every crossing set of size greater than $\beta P$.

Choose a closed box containing, in its interior, a representative of every realizable sign pattern of the affine functions. Such a box exists because there are finitely many sign patterns. Consider the arrangement of the hyperplanes corresponding to the selected nonconstant affine functions within this box. Triangulate its cells and retain the relative interiors of the resulting simplices of all dimensions. The number of simplices depends only on $a$ and the number of selected hyperplanes. Moreover, every selected affine function has a constant sign on the relative interior of each simplex.

Suppose that the predicate $\ind[\ell_j(z)\ne0]$ is nonconstant on the relative interior of a simplex with vertices $q_0,\ldots,q_t$. There is a point $z$ in this relative interior such that $\ell_j(z)=0$, and another point at which $\ell_j$ is nonzero. Write
\[
 z=\sum_{\nu=0}^t\alpha_\nu q_\nu,\qquad
 \alpha_\nu>0,\qquad \sum_{\nu=0}^t\alpha_\nu=1.
\]
Since $\ell_j$ is affine,
\[
 0=\ell_j(z)=\sum_{\nu=0}^t\alpha_\nu\ell_j(q_\nu).
\]
If all vertex values were nonnegative or all were nonpositive, they would all have to be zero. The function would then vanish throughout the simplex, contradicting the assumption that the predicate is nonconstant. Thus $\ell_j$ takes both strictly positive and strictly negative values at the vertices.

No selected affine function can have strictly opposite signs at the vertices of a simplex in this triangulation, since that would also give opposite signs in its relative interior. Consequently, at most $\beta P$ predicates can be nonconstant on the relative interior of any simplex: otherwise, its crossing set would contain one of the selected indices. For each simplex, take the coordinatewise minimum and maximum of the binary vectors realized on its relative interior. These two vectors define a bracket of width at most $\beta$. The resulting brackets cover every realizable binary vector, including those attained on cells of lower dimension. Predicates that are identically zero do not contribute to the bracket widths.

\par\smallskip\noindent\textbf{Pairwise independent samples.}\enspace
Choose $v$ distinct elements $t_1,\ldots,t_v\in\mathbb F_P$. For every pair of seeds $a_0,b_0\in\mathbb F_P$, include the ordered sample
\[
 (a_0t_1+b_0,\ldots,a_0t_v+b_0)
\]
in $\mathcal F$. Thus the family contains $P^2$ samples and can be enumerated without knowing the affine functions.

For uniformly and independently chosen seeds $a_0,b_0$, the sampled indices are uniform and pairwise independent over $\mathbb F_P$. For any fixed binary vector $w\in\{0,1\}^P$, the sample mean $v^{-1}\sum_{j\in J}w_j$ has expectation $P^{-1}\sum_jw_j$ and variance at most $1/(4v)$.

Set $\beta=\eta/4$ and consider the finite collection of brackets constructed above. By Chebyshev's inequality and a union bound over their endpoints, we can choose $v$, depending only on $a,\eta$, so that with probability at least $1-\eta/2$, every bracket endpoint has sample error at most $\eta/4$.

On this event, consider any realizable vector $h(z)$ and a bracket $[l,u]$ containing it. The coordinatewise inequalities $l\le h(z)\le u$ hold both for the population averages and for the sample averages. Since the width of the bracket is at most $\beta$, the sample error of $h(z)$ is at most
\[
 \beta+\eta/4=\eta/2.
\]
Thus the maximum sample error over all $z\in\R^a$ is at most $\eta/2$ with probability at least $1-\eta/2$. On the complementary event, the error is at most one. Its expectation is therefore at most $\eta$.

We may choose $v\ge2$. Bertrand's postulate guarantees a prime $P$ in the required interval for $N\ge v$, and such a prime can be found by trial division in polynomial time in $N$. The bracket construction is used only in the analysis; the algorithm enumerates the samples directly from the two seeds.
\end{proof}

\subsection{Deterministic additive approximation}\label{sec:detadd}

\begin{figure}[!htbp]
\centering
\begin{tikzpicture}[emvdfig]
  \node[emvdhead] at (.12,0) {Process the blocks in order};
  \draw[emvdBlue,line width=.85pt] (.12,-.25)--(14.99,-.25);
  \foreach \x/\cl in {.20/emvdBlue,3.20/emvdBlue,6.20/emvdTeal,9.20/emvdAmber,12.20/emvdAmber}{
    \draw[rounded corners=1mm,draw=\cl!70,fill=\cl!7] (\x,-.62) rectangle ++(2.70,-.88);
    \foreach \dx/\dy in {.38/.22,1.00/.60,1.65/.26,2.29/.56}
      \node[circle,fill=\cl,inner sep=0pt,minimum size=4.2pt] at ({\x+\dx},{-.62-\dy}) {};
  }
  \node[emvdnote] at (3.05,-1.76) {Earlier blocks $P_t$};
  \node[emvdnote] at (7.55,-1.76) {Current block $B_t$};
  \node[emvdnote] at (12.05,-1.76) {Later blocks $Q_t$};
  \draw[emvdarrow] (3.05,-1.96)--(3.05,-2.28);
  \draw[emvdarrow] (12.05,-1.96)--(12.05,-2.28);
  \node[draw=emvdBlue!60,fill=emvdBlue!5,rounded corners=1mm,text width=5.25cm,minimum height=1.25cm,align=center,font=\small] (prev) at (3.05,-3.02)
    {Compress known loss predicates\\$\ell_{ij}(x,y_j(\boldsymbol{\xi}))$};
  \node[draw=emvdAmber!70,fill=emvdAmber!5,rounded corners=1mm,text width=5.25cm,minimum height=1.25cm,align=center,font=\small] (next) at (12.05,-3.02)
    {Fixed sample of later indices\\with unknown positions in $\boldsymbol{\xi}$};
  \node[draw=emvdTeal!70,fill=emvdTeal!6,rounded corners=1mm,text width=6.0cm,minimum height=.98cm,align=center,font=\small] (local) at (7.55,-4.65)
    {Weighted local minimization\\$y_i(\boldsymbol{\xi})$, for each $i\in B_t$};
  \draw[emvdarrow] (prev.south) |- (local.west);
  \draw[emvdarrow] (next.south) |- (local.east);
  \draw[emvdarrow] (local.south)--(7.55,-5.44);
  \node[emvdresult] at (7.55,-5.86)
    {After all blocks: compress the pair losses, optimize $\boldsymbol{\xi}$, and score the result on all pairs.};
\end{tikzpicture}
\caption{Block processing for one sample from \cref{lem:universal}. The earlier placements are known functions of $\boldsymbol{\xi}$, and the later reference positions are represented by variables. Each local minimization uses both sources. After processing all blocks, the algorithm chooses $\boldsymbol{\xi}$ and evaluates the resulting configuration on all pairs.}
\label{fig:additive-blocks}
\end{figure}
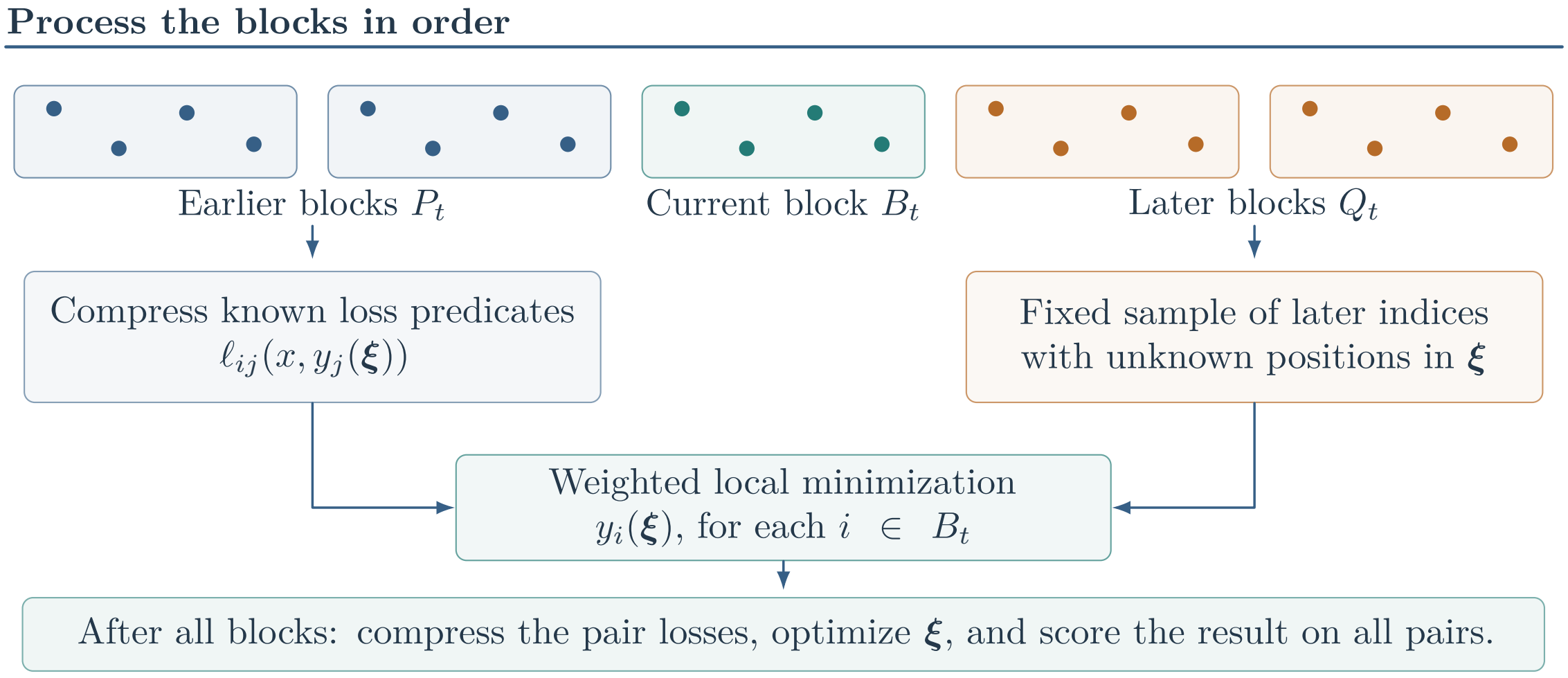

We process the points in blocks. For each fixed sample of indices from \cref{lem:universal}, introduce a vector of real variables for the unknown optimal reference positions. The earlier placements become functions of this vector, so their loss predicates can be compressed using \cref{lem:detcompress}. A final compression of the pairwise losses allows us to choose the variables; see \cref{fig:additive-blocks}.

\FloatBarrier
\Needspace{9\baselineskip}
\begin{lemma}[Deterministic additive approximation]\label{lem:detadd}
For either geometric problem in dimension $p$ and any $0<\delta\le1$, there is a deterministic algorithm that returns a feasible configuration of cost at most $k+\delta n^2$ in time
\[
 f(p,\delta)(n+L)^{O(1)},
\]
where $f$ is a computable function and the polynomial exponent is absolute. The same guarantee holds for algebraic input values of bounded individual degrees, with the degree bound included among the parameters. Moreover, if $B$ bounds the individual input encoding lengths, every output coordinate has algebraic degree bounded in terms of the parameters and encoding length at most $f(p,\delta)(B+1+\log(n+1))$, with the degree bound also included in $f$ when necessary. Here $L$ denotes the total encoding length of the current instance.
\end{lemma}

\begin{proof}
If $p=0$ or $n\le1$, we solve the instance exactly as described in \cref{sec:prelim}. Otherwise, partition $V$ into
\[
 b=\min\{n,\lceil4/\delta\rceil\}
\]
blocks $B_1,\ldots,B_b$ whose sizes differ by at most one. For each block $B_t$, let
\[
 P_t=\bigcup_{s<t}B_s,\qquad Q_t=\bigcup_{s>t}B_s
\]
denote the indices in the preceding and following blocks, respectively.

\par\smallskip\noindent\textbf{Constructing a parameterized configuration.}\enspace
Apply \cref{lem:universal} with $a=p+1$ and accuracy $\eta_0=\delta/128$, and let $v$ be the resulting sample length. If $n<v$, the number of points is bounded in terms of the parameters, and we solve the instance exactly. Otherwise, let $P_0\in[n,2n)$ be the prime used in the lemma and let $\mathcal F$ be the corresponding family of $P_0^2$ ordered samples.

For each sample $(q_1,\ldots,q_v)\in\mathcal F$, introduce a vector $\xi_h\in\R^p$ for every occurrence $h\in[v]$ with $q_h\in V$, representing its unknown comparison position. Let $\boldsymbol{\xi}$ collect these variables (at most $pv$ coordinates). The same sample is used in all blocks; dummy indices need no variables.

Suppose we have already constructed the positions $y_j(\boldsymbol{\xi})$ for all $j\in P_t$. For each $i\in B_t$, if $P_t$ is nonempty, apply \cref{lem:detcompress} with accuracy $\delta/128$ to the predicates
\[
 (x,\boldsymbol{\xi})\longmapsto\ell_{ij}(x,y_j(\boldsymbol{\xi})),\qquad j\in P_t.
\]
Let $J_i$ be the resulting multiset of reference indices. Its size depends only on $p$ and $\delta$. The compression guarantees that
\[
 \left|\sum_{j\in P_t}\ell_{ij}(x,y_j(\boldsymbol{\xi}))
 -\frac{|P_t|}{|J_i|}\sum_{j\in J_i}\ell_{ij}(x,y_j(\boldsymbol{\xi}))\right|
 \le\frac{\delta n}{128}
\]
for every $x$ and $\boldsymbol{\xi}$. When $P_t$ is empty, no compression is needed, and the estimated cost to the preceding blocks is zero.

For the following blocks, we use the occurrences of the fixed sample whose indices belong to $Q_t$. Define the estimated local cost of placing point $i$ at $x\in\mathcal D_i$ by
\begin{align*}
 \widehat C_i(x;\boldsymbol{\xi})={}&
 \frac{|P_t|}{|J_i|}\sum_{j\in J_i}\ell_{ij}(x,y_j(\boldsymbol{\xi}))\\
 &+\frac{P_0}{v}\sum_{\substack{h\in[v]\\q_h\in Q_t}}\ell_{i q_h}(x,\xi_h),
\end{align*}
where the first term is defined to be zero when $P_t$ is empty. Both sums count sample occurrences with multiplicity. The factor $P_0/v$ converts the sample average into an estimate of the total cost to the following blocks.

Define $y_i(\boldsymbol{\xi})$ as an optimal local placement for this weighted objective, with the canonical selection rule of \cref{lem:local}. Inductively, each placement depends on a number of previous placements bounded in terms of $p,\delta$, with dependency depth at most $b$. Bounded-size quantifier elimination therefore gives descriptions of the placements and their loss predicates whose complexity depends only on $p,\delta$. All samples and weights are fixed independently of $\boldsymbol{\xi}$.

The configuration $Y(\boldsymbol{\xi})=(y_i(\boldsymbol{\xi}))_{i\in V}$ is feasible for every $\boldsymbol{\xi}$, since local minimization is restricted to $\mathcal D_i$ even if a sampled reference position is infeasible.

\par\smallskip\noindent\textbf{Bounding the additive error.}\enspace
Fix an optimal configuration $X^*=(x_i^*)_{i\in V}$. For a given sample, let $\boldsymbol{\xi}^*$ assign $\xi_h=x_{q_h}^*$ to every occurrence with $q_h\in V$.

For each $i\in B_t$, define
\[
 e_i=\sup_{x\in\mathcal D_i}\left|
 \frac1v\sum_{\substack{h\in[v]\\q_h\in Q_t}}\ell_{i q_h}(x,x_{q_h}^*)
 -\frac1{P_0}\sum_{j\in Q_t}\ell_{ij}(x,x_j^*)\right|.
\]
The predicates for indices outside $Q_t$, including dummies, are zero.

For EMVD, the loss $\ell_{ij}(x,x_j^*)$ tests whether
\[
 \|x\|^2-2\langle x,x_j^*\rangle+\|x_j^*\|^2-A_{ij}\ne0.
\]
This is a test of nonzero value of an affine function in the lifted variables $z=(x,\|x\|^2)\in\R^{p+1}$. For the prescribed-norm problem, the loss tests whether
\[
 \langle x,x_j^*\rangle-g_{ij}\ne0,
\]
which is already an affine test in $x$ and can be viewed as a test in the same lifted variables. Thus \cref{lem:universal} gives
\[
 \E_{J\sim\mathcal F}[e_i]\le\frac{\delta}{128}.
\]
By linearity of expectation, there exists a sample in $\mathcal F$ satisfying
\[
 \sum_{i\in V}e_i\le\frac{\delta n}{128}.
\]
Fix such a sample for the remainder of the analysis.

Let $F_t$ denote the cost of the configuration in which the points in the first $t$ blocks are placed at $y_i(\boldsymbol{\xi}^*)$ and all other points remain at their optimal positions $x_i^*$. In particular,
\[
 F_0=k,\qquad F_b=\err(Y(\boldsymbol{\xi}^*)).
\]
For a point $i\in B_t$, define its true local cost with respect to points outside the current block by
\begin{align*}
 C_i(x)={}&\sum_{j\in P_t}\ell_{ij}(x,y_j(\boldsymbol{\xi}^*))\\
 &+\sum_{j\in Q_t}\ell_{ij}(x,x_j^*).
\end{align*}
The guarantees of the two sampling lemmas imply that
\[
 |\widehat C_i(x;\boldsymbol{\xi}^*)-C_i(x)|\le P_0e_i+\frac{\delta n}{128}
\]
for every $x\in\mathcal D_i$. Since $y_i(\boldsymbol{\xi}^*)$ minimizes the estimated local cost, we have
\begin{align*}
 C_i(y_i(\boldsymbol{\xi}^*))
 &\le \widehat C_i(y_i(\boldsymbol{\xi}^*);\boldsymbol{\xi}^*)+P_0e_i+\frac{\delta n}{128}\\
 &\le \widehat C_i(x_i^*;\boldsymbol{\xi}^*)+P_0e_i+\frac{\delta n}{128}\\
 &\le C_i(x_i^*)+2\left(P_0e_i+\frac{\delta n}{128}\right).
\end{align*}
Accounting also for pairs within $B_t$, we obtain
\[
 F_t-F_{t-1}\le 2\sum_{i\in B_t}\left(P_0e_i+\frac{\delta n}{128}\right)+\binom{|B_t|}{2}.
\]
If $b=n$, every block contains a single point and there are no internal pairs. Otherwise, $b\ge4/\delta$ and $|B_t|-1\le n/b$ for every $t$. Hence
\[
 \sum_{t=1}^b\binom{|B_t|}{2}\le\frac{n^2}{2b}\le\frac{\delta n^2}{8}.
\]
Summing over all blocks and using $P_0<2n$ gives
\begin{align*}
 \err(Y(\boldsymbol{\xi}^*))&=F_b\\
 &\le k+2P_0\sum_{i\in V}e_i+\frac{\delta n^2}{64}+\frac{\delta n^2}{8}\\
 &\le k+\frac{11\delta n^2}{64}.
\end{align*}
The later reference losses depend only on the fixed $X^*$, while compression of the earlier reference losses is uniform in $(x,\boldsymbol{\xi})$. Thus the adaptive dependence of earlier placements on the sample does not affect the error bounds.

\par\smallskip\noindent\textbf{Selecting the parameter vector.}\enspace
To choose the parameters without knowing $X^*$, compress the total pairwise loss for each sample.

For each sample in $\mathcal F$, consider the predicates
\[
 \boldsymbol{\xi}\longmapsto\ell_{ij}(y_i(\boldsymbol{\xi}),y_j(\boldsymbol{\xi})),\qquad 1\le i<j\le n.
\]
These predicates have complexity depending only on $p,\delta$. Apply \cref{lem:detcompress} with accuracy $\delta/8$ to obtain a multiset $J_{\mathrm{pair}}$ of unordered pairs. Define the estimated total cost
\[
 \widehat F(\boldsymbol{\xi})=\frac{\binom{n}{2}}{|J_{\mathrm{pair}}|}
 \sum_{\{i,j\}\in J_{\mathrm{pair}}}\ell_{ij}(y_i(\boldsymbol{\xi}),y_j(\boldsymbol{\xi})).
\]
For every $\boldsymbol{\xi}$,
\[
 |\widehat F(\boldsymbol{\xi})-\err(Y(\boldsymbol{\xi}))|\le\frac{\delta}{8}\binom{n}{2}.
\]
Minimize $\widehat F$ exactly using \cref{sec:prelim}.

Let $\widehat{\boldsymbol{\xi}}$ be a minimizer. On the sample satisfying the error bound above,
\begin{align*}
 \err(Y(\widehat{\boldsymbol{\xi}}))
 &\le \widehat F(\widehat{\boldsymbol{\xi}})+\frac{\delta}{8}\binom{n}{2}\\
 &\le \widehat F(\boldsymbol{\xi}^*)+\frac{\delta}{8}\binom{n}{2}\\
 &\le \err(Y(\boldsymbol{\xi}^*))+\frac{\delta}{4}\binom{n}{2}\\
 &\le k+\frac{19\delta n^2}{64}\le k+\delta n^2.
\end{align*}
The algorithm constructs one configuration for each sample in $\mathcal F$, evaluates its cost on all unordered pairs, and returns the configuration of minimum cost.

\par\smallskip\noindent\textbf{Running time and algebraic complexity.}\enspace
There are $O(n^2)$ samples. Each requires at most one predicate compression and one symbolic local minimization per point, followed by compression of the pair losses and optimization over $\boldsymbol{\xi}$. All descriptions involved in local minimization have complexity depending only on $p,\delta$. The weights have bit length $O_{p,\delta}(1+\log(n+1))$. Each description depends on a bounded number of numerical inputs and has bounded dependency depth. By \cref{sec:prelim}, output coordinates have bounded individual algebraic degrees and encoding length at most $f(p,\delta)(B+1+\log(n+1))$, including the input degree bound in $f$. The same bound applies when evaluating a placement at the selected parameter values: the compressed final objective involves only a bounded number of such descriptions. Compression returns only index multisets, not algebraic values from its internal sign-pattern computations. Exact cost evaluation uses a bounded number of algebraic values per pair. By \cref{lem:detcompress}, the total time is $f(p,\delta)(n+L)^{O(1)}$ with an absolute exponent.
\end{proof}

\subsection{Concentrated subspaces and voting}\label{sec:concentration}

We next establish the geometric properties needed to turn the additive guarantee into a relative guarantee for instances with few disagreements.

\paragraph{Concentration and separation.}
Fix a feasible comparison configuration $X^*=(x_i^*)_{i\in V}$ in dimension $p$. For this subsection, let $k=\err(X^*)$ and define the error degree of point $i$ by
\[
 b_i=\sum_{j\in V\setminus\{i\}}\ell_{ij}(x_i^*,x_j^*).
\]
Thus $\sum_{i\in V}b_i=2k$. Use the thresholds defined in \eqref{eq:threshold}, and assume $n\ge4/\theta$, since smaller instances can be solved exactly. In the prescribed-norm problem, assume that points of zero norm have already been removed.

Let $r$ be the smallest dimension for which there exists a subspace containing at least $(1-\tau_r)n$ points of $X^*$. Here and below, subspaces are affine for EMVD and linear for the prescribed-norm problem. Such a dimension exists because $\tau_p=0$. Fix one such subspace $L_0$ and define
\[
 C=\{i\in V:x_i^*\in L_0\},\qquad \lambda=\tau_{r-1}-\tau_r.
\]
By definition, $|C|\ge(1-\tau_r)n$. If $r\ge1$, the minimality of $r$ implies that every proper subspace of $L_0$ of the appropriate type contains fewer than $(1-\tau_{r-1})n$ points of $X^*$. Consequently, every such proper subspace excludes at least $\lambda n$ points indexed by $C$.

If $r=0$ in EMVD, all points indexed by $C$ coincide and $|C|\ge\lambda n$. In the prescribed-norm problem, we have already removed points of zero norm, so $r\ge1$. The thresholds in \eqref{eq:threshold} give
\begin{equation}\label{eq:separate}
 \lambda\ge\theta,\qquad |V\setminus C|\le\tau_r n,\qquad \frac{\tau_r}{\lambda}\le\frac{\eps}{2048}.
\end{equation}
Thus $L_0$ contains most of the comparison points, while its proper subspaces exclude substantially more points than there are outside $L_0$.

\paragraph{Reliable anchors.}
We need anchors that span $L_0$, agree with their mutual observations, and have few erroneous incident observations.

\begin{lemma}[Reliable anchors]\label{lem:anchors}
Suppose $k\le\gamma n^2$. There exist $r+1$ affinely independent points indexed by $C$ for EMVD, or $r$ linearly independent points indexed by $C$ for the prescribed-norm problem, such that all mutual observations between these points agree with $X^*$. Moreover, their total error degree is at most
\[
 \frac{4(p+1)k}{\lambda n}.
\]
\end{lemma}

\begin{proof}
Since $\sum_i b_i=2k$, at most $\lambda n/2$ points have error degree greater than $4k/(\lambda n)$. When $k=0$, every point has error degree zero.

We select the anchors one at a time, extending their affine or linear span at each step. Until the selected points span $L_0$, there are at least $\lambda n$ points indexed by $C$ outside their current span. For the first affine anchor, all points indexed by $C$ are eligible.

Exclude points with error degree greater than $4k/(\lambda n)$ and points having an incorrect observation to any previously selected anchor. Since every previously selected anchor has error degree at most $4k/(\lambda n)$, the number of points excluded by the second condition is at most
\[
 \frac{4(p+1)k}{\lambda n}\le\frac{4(p+1)\gamma n}{\lambda}\le\frac{\lambda n}{8},
\]
where the last inequality follows from the definition of $\gamma$ and the fact that $\lambda\ge\theta$.

At least $3\lambda n/8$ eligible points remain, so the selection can continue until the anchors span $L_0$. By construction, all mutual observations between the anchors are correct. Summing their error degrees gives the claimed bound.
\end{proof}

\paragraph{Recovering projections from anchors.}
For EMVD, let $a_0,\ldots,a_r$ be the anchor indices. Choose coordinates on $L_0$ so that the corresponding anchor positions are $v_0=0,v_1,\ldots,v_r\in\R^r$. For any point $i$ whose observations to these anchors are correct, let $u_i\in\R^r$ be its projection coordinates and let $R_i$ be its squared distance to $L_0$. The anchor observations give
\[
 A_{a_ji}=\|u_i-v_j\|^2+R_i,\qquad 0\le j\le r.
\]
Subtracting the equation for anchor $a_j$ from the equation for anchor $a_0$ yields
\begin{equation}\label{eq:projection}
 \begin{aligned}
 2\langle u_i,v_j\rangle&=\|v_j\|^2+A_{a_0i}-A_{a_ji},&&1\le j\le r,\\
 R_i&=A_{a_0i}-\|u_i\|^2.
 \end{aligned}
\end{equation}
Since $v_1,\ldots,v_r$ are linearly independent, these equations uniquely determine $u_i$ and $R_i$. When $r=0$, only the squared distance $R_i$ remains to be determined.

For the prescribed-norm problem, let $a_1,\ldots,a_r$ be the anchor indices, with linearly independent positions $v_1,\ldots,v_r$. The corresponding equations are
\[
 \langle u_i,v_j\rangle=g_{ia_j},\qquad 1\le j\le r,
\]
and
\[
 R_i=s_i-\|u_i\|^2.
\]
Again, the projection and squared distance to $L_0$ are uniquely determined.

\paragraph{Voting with inaccurate references.}
We next consider the case where some reference positions are inaccurate. Let $T\subseteq V$ be a set of reference indices, with positions $v_j\in\R^r$ for $j\in T$. After identifying $L_0$ with $\R^r$ using an orthonormal coordinate frame, let $v_j^*$ denote the true reference position of each $j\in C$.

Assume that the two sets of reference data differ on at most $b$ indices, that is,
\[
 |T\mathbin{\triangle}C|+|\{j\in T\cap C:v_j\ne v_j^*\}|\le b.
\]
Thus a discrepancy is a missing reference, an extra reference, or an incorrect position.

To compute a feasible pair satisfying \eqref{eq:approxvote}, apply \cref{lem:detcompress} with accuracy $\eta/2$ to the loss predicates associated with the reference indices in $T\setminus\{i\}$. Give each sampled occurrence weight
\[
 \frac{|T\setminus\{i\}|}{|J_i|},
\]
where $J_i$ is the resulting multiset. The weighted objective differs from $c_i(u,R)$ by at most $\eta n/2$, uniformly over all feasible pairs $(u,R)$. Minimizing the weighted objective exactly using the algebraic tools of \cref{sec:prelim} therefore gives a feasible pair $(\widehat u_i,\widehat R_i)$ satisfying \eqref{eq:approxvote}. If $T\setminus\{i\}$ is empty, we choose any feasible pair.

\begin{lemma}[Stable voting]\label{lem:vote}
Suppose $b\le\lambda n/8$ and $\eta<\lambda/4$. Let
\[
 H=\{i\in V:b_i>\lambda n/8\},\qquad h=|H|.
\]
Then $h\le16k/(\lambda n)$. For every $i\notin H$, the pair $(\widehat u_i,\widehat R_i)$ selected by \eqref{eq:approxvote} equals the true projection coordinates and squared distance of $x_i^*$ to $L_0$. Moreover, the selected pairs admit a feasible completion of cost at most
\begin{equation}\label{eq:votebound}
 k+2hb+3h^2+h\tau_r n+\eta nh.
\end{equation}
\end{lemma}

\begin{proof}
The bound on $h$ follows immediately from $\sum_i b_i=2k$.

\par\smallskip\noindent\textbf{Exact recovery for points of small error degree.}\enspace
For EMVD, consider two distinct feasible pairs $(u,R)$ and $(u',R')$. Their predicted squared distances to a reference position $v\in\R^r$ agree if and only if
\[
 \|u-v\|^2+R=\|u'-v\|^2+R',
\]
or equivalently,
\[
 2\langle u-u',v\rangle=\|u\|^2+R-\|u'\|^2-R'.
\]
If $u\ne u'$, this equation defines a proper affine hyperplane. If $u=u'$, then $R\ne R'$ and the equation has no solution.

For the prescribed-norm problem, the squared distance $R$ is determined by $u$ and $s_i$. Hence two distinct feasible pairs have different projections, and their predicted inner products agree only when
\[
 \langle u-u',v\rangle=0.
\]
This equation defines a proper linear hyperplane.

By the separation property of $L_0$, any two distinct feasible pairs therefore give different predictions on at least $\lambda n$ of the true reference points. This also holds when $r=0$ in EMVD, since distinct feasible pairs have different constant squared distance predictions.

Fix a point $i\notin H$ and let $(u_i^*,R_i^*)$ be its true projection coordinates and squared distance. Among the true references indexed by $C\setminus\{i\}$, the true pair disagrees with at most $b_i$ observations. Any other feasible pair disagrees with at least $\lambda n-1-b_i$ observations. The subtraction of one accounts for the exclusion of point $i$ itself.

Replacing the true reference data by the reference data on $T$ changes the voting cost of any feasible pair by at most $b$. Thus, for every feasible pair $(u,R)\ne(u_i^*,R_i^*)$,
\begin{align*}
 c_i(u,R)-c_i(u_i^*,R_i^*)
 &\ge\lambda n-1-2b_i-2b\\
 &\ge\lambda n/2-1\\
 &\ge\lambda n/4\\
 &>\eta n.
\end{align*}
Here we used $b_i\le\lambda n/8$, $b\le\lambda n/8$, and $n\ge4/\theta\ge4/\lambda$. Therefore, no other feasible pair satisfies \eqref{eq:approxvote}, and the selected pair must equal $(u_i^*,R_i^*)$.

\par\smallskip\noindent\textbf{Bounding the completion cost.}\enspace
Construct a comparison completion by keeping the positions $x_i^*$ for every $i\notin H$ and choosing any feasible completion of $(\widehat u_i,\widehat R_i)$ for each $i\in H$. Such a completion exists by the feasibility conditions.

Let $C_0=C\setminus H$. Every point indexed by $C_0$ retains its true position in $L_0$. For a point $i\in H$, its distances or inner products to points indexed by $C_0$ depend only on its projection and squared distance to $L_0$, and not on the direction of its residual vector.

The reference data on $T$ differ from the true reference data on $C_0$ on at most $b+h$ indices. Consequently, replacing one reference set by the other changes the local cost of any feasible pair by at most $b+h$. By \eqref{eq:approxvote}, changing the true pair of point $i$ to the selected pair increases the number of disagreements between $i$ and $C_0$ by at most
\[
 2(b+h)+\eta n.
\]
Summing over $i\in H$, these pairs contribute at most $2h(b+h)+\eta nh$ additional disagreements.

All other pairs whose cost can increase have one endpoint in $H$ and the other outside $C_0$. Since
\[
 |V\setminus C_0|\le |V\setminus C|+h\le\tau_r n+h,
\]
there are at most $h(\tau_r n+h)$ such pairs. Each contributes at most one additional disagreement.

Pairs with both endpoints outside $H$ retain their original positions and hence their original costs. The total cost of the comparison completion is therefore at most
\begin{align*}
 k+2h(b+h)+\eta nh+h(\tau_r n+h)
 &=k+2hb+3h^2+h\tau_r n+\eta nh,
\end{align*}
as claimed.
\end{proof}

\paragraph{The residual problem.}
Once the pairs $(\widehat u_i,\widehat R_i)$ have been selected, write each completed point in the coordinates adapted to the subspace as
\[
 (\widehat u_i,z_i),\qquad z_i\in\R^{p-r},\qquad \|z_i\|^2=\widehat R_i.
\]
For EMVD, orthogonality gives
\begin{align*}
 \|x_i-x_j\|^2&=\|\widehat u_i-\widehat u_j\|^2+\|z_i-z_j\|^2\\
 &=\|\widehat u_i-\widehat u_j\|^2+\widehat R_i+\widehat R_j-2\langle z_i,z_j\rangle.
\end{align*}
Thus the original distance constraint is satisfied if and only if
\[
 \langle z_i,z_j\rangle=\frac{\|\widehat u_i-\widehat u_j\|^2+\widehat R_i+\widehat R_j-A_{ij}}{2}.
\]
For the prescribed-norm problem, the subspace is linear, and
\[
 \langle x_i,x_j\rangle=\langle\widehat u_i,\widehat u_j\rangle+\langle z_i,z_j\rangle.
\]
Hence the original inner product constraint is satisfied if and only if
\[
 \langle z_i,z_j\rangle=g_{ij}-\langle\widehat u_i,\widehat u_j\rangle.
\]
These are precisely the residual constraints defined in \eqref{eq:residual}.

If $\widehat R_i=0$, then $z_i=0$, and all constraints involving $z_i$ have fixed costs. The completion cost is therefore the sum of a fixed nonnegative cost and the objective of a prescribed-norm instance in dimension $p-r$. If $r=p$, every residual vector is zero and no optimization remains. Otherwise, we solve the residual instance recursively.

\subsection{Recursive reconstruction and proof of the EPTAS}\label{sec:detmult}

We now prove the approximation guarantee and running time of \cref{alg:eptas}. The main step is to show that, when the optimum is small, the candidate subspaces generated from the additive solution include one whose reference positions satisfy the conditions of \cref{lem:vote}.

\begin{proof}[Proof of \cref{thm:schemes}]
We first prove the approximation guarantee for the prescribed-norm problem by induction on the dimension $p$, for every accuracy $0<\eps\le1$. We then prove the guarantee for EMVD using the result for the prescribed-norm problem.

In the prescribed-norm problem, remove all points of zero norm and account separately for the fixed cost of pairs involving these points. Since this cost is nonnegative, a multiplicative approximation for the remaining instance gives the same approximation ratio for the original instance. In what follows, $V,n,k$ refer to the remaining instance.

Instances in dimension zero or with $n\le1$ can be solved directly. Use the parameters defined in \eqref{eq:threshold} and \eqref{eq:eptas-parameters}. If $n<4/\theta$, solve the instance exactly as described in \cref{sec:prelim}. Otherwise, compute a configuration $Y=(y_i)_{i\in V}$ using \cref{lem:detadd} with accuracy $\delta$. Its cost satisfies
\[
 \err(Y)\le k+\delta n^2.
\]
If $k\ge\gamma'n^2$, then, since $\delta=\eps\gamma'$,
\[
 \err(Y)\le k+\eps\gamma'n^2\le(1+\eps)k.
\]
The algorithm retains $Y$ among its candidate outputs, so it suffices to consider the case $k\le\gamma'n^2$.

\par\smallskip\noindent\textbf{Finding a reliable subspace in the additive solution.}\enspace
Fix an optimum $X^*$ and its subspace $L_0$, with $r,C,\lambda$ as in \cref{sec:concentration}.

Let $k'$ denote the number of pairs for which the squared distances or inner products induced by $X^*$ and $Y$ are different. For every such pair, at least one of the two configurations disagrees with the original observation. Consequently,
\[
 \begin{aligned}
 k'&\le\err(X^*)+\err(Y)\\
 &\le 2k+\delta n^2\\
 &\le3\gamma'n^2.
 \end{aligned}
\]
In the prescribed-norm problem, both configurations satisfy the same norm constraints.

Apply \cref{lem:anchors}, treating the squared distances or inner products induced by $Y$ as the observations and $X^*$ as the comparison configuration. Since $3\gamma'\le\gamma$, the lemma guarantees a set of anchors in $C$ spanning $L_0$ whose mutual observations agree in the two configurations. Let $b_0$ be the number of indices whose observations to at least one of these anchors differ between $X^*$ and $Y$. The sum of the anchor error degrees bounds $b_0$, and hence
\begin{equation}\label{eq:detB}
 \begin{aligned}
 b_0&\le\frac{4(p+1)k'}{\lambda n}\\
 &\le\frac{12(p+1)\gamma'n}{\lambda}\\
 &\le\frac{\eps\lambda n}{4096}.
 \end{aligned}
\end{equation}
The last inequality uses $\lambda\ge\theta$.

Let $L_1$ be the affine or linear span of the corresponding anchor positions in $Y$. Since the mutual squared distances or inner products of the anchors agree, the two anchor configurations are congruent (in the inner product case, the diagonal entries also agree). We may therefore align $X^*$ with $Y$ by a Euclidean isometry for EMVD or an orthogonal transformation for the prescribed-norm problem, so that the corresponding anchor positions coincide and $L_0$ is mapped to $L_1$. We continue to denote the aligned comparison configuration by $X^*$.

By \eqref{eq:projection} and its inner product analogue, every point whose observations to all the anchors agree has the same projection onto $L_1$ and the same squared distance to $L_1$ in the two aligned configurations. In particular, for every such point $i$,
\[
 y_i\in L_1\quad\Longleftrightarrow\quad x_i^*\in L_1,
\]
and if these conditions hold, then $y_i=x_i^*$. It follows that the reference data consisting of all points of $Y$ in $L_1$ differ from the true reference data indexed by $C$ on at most $b_0$ indices.

\par\smallskip\noindent\textbf{Enumerating candidate subspaces.}\enspace
The membership and difference predicates used in \cref{alg:eptas} have complexity bounded in terms of $p$: a subspace can be specified by at most $p$ affine or homogeneous linear equations, with redundant equations allowed. Thus \cref{lem:detcompress} produces the two samples at accuracy $\theta/4$, whose union of indices $Q$ has size depending only on $p,\eps$.

By the separation property of $L_0$ and the reference discrepancy bound \eqref{eq:detB}, every proper subspace $L'\subsetneq L_1$ of the appropriate type satisfies
\[
 |\{i:y_i\in L_1\setminus L'\}|\ge\lambda n-b_0.
\]
Since
\[
 \lambda-\frac{b_0}{n}>\frac{\theta}{4},
\]
the sample obtained from the difference predicates must contain an index $i$ with $y_i\in L_1\setminus L'$ for every such $L'$. Similarly,
\[
 |\{i:y_i\in L_1\}|\ge |C|-b_0\ge\lambda n-b_0,
\]
so the sample obtained from the membership predicates contains at least one point in $L_1$.

The sampled points in $L_1$ must span it: otherwise their span is a proper subspace $L'$ with no sampled point in $L_1\setminus L'$. The membership sample covers the zero-dimensional affine case as well. A spanning tuple of at most $p+1$ points for EMVD, or at most $p$ points for the prescribed-norm problem, is therefore enumerated by \cref{alg:eptas}.

\par\smallskip\noindent\textbf{Voting and recursive completion.}\enspace
Consider the candidate $S=L_1$. Its reference positions differ from the true reference positions on at most $b_0$ indices. By \eqref{eq:detB} and the definition of $\eta$,
\[
 b_0\le\lambda n/8,\qquad \eta<\lambda/4.
\]
Hence the conditions of \cref{lem:vote} hold, using the original observations of the current instance.

Let $h$ be the number of points with error degree greater than $\lambda n/8$ in $X^*$ relative to these original observations. By \cref{lem:vote},
\[
 h\le\frac{16k}{\lambda n}.
\]
The projections and squared distances selected by voting admit a feasible completion of cost at most
\[
 k+2hb_0+3h^2+h\tau_r n+\eta nh.
\]
Using \eqref{eq:separate}, \eqref{eq:detB}, $k\le\gamma'n^2$, and the parameter definitions, we obtain
\begin{align*}
 2hb_0&\le\frac{\eps k}{128},\\
 3h^2&\le\frac{768\gamma'}{\lambda^2}k,\\
 h\tau_r n&\le\frac{\eps k}{128},\\
 \eta nh&\le\frac{\eps k}{32}.
\end{align*}
Since
\[
 \frac{768\gamma'}{\lambda^2}\le\frac{3\eps}{256(p+1)}\le\frac{3\eps}{256},
\]
the total additional cost is at most
\[
 \left(\frac{\eps}{128}+\frac{3\eps}{256}+\frac{\eps}{128}+\frac{\eps}{32}\right)k
 =\frac{15\eps k}{256}\le\frac{\eps k}{4}.
\]
Thus the minimum cost of a completion consistent with the selected projections and squared distances is at most $(1+\eps/4)k$.

Once these quantities are fixed, the remaining optimization is the prescribed-norm instance $I_S$ defined in \eqref{eq:residual}. Let $c_S$ denote the fixed cost of the pairs involving zero residual vectors, and let $k_S$ denote the optimum cost of the residual instance restricted to points of positive residual norm, taking $k_S=0$ when there are no such points. By the equivalence established in \cref{sec:concentration},
\[
 c_S+k_S\le(1+\eps/4)k.
\]
If $r=p$, all residual vectors are zero and the completion requires no further optimization. Otherwise, the algorithm solves $I_S$ recursively with accuracy $\eps/4$.

For the prescribed-norm problem, every enumerated subspace has positive dimension, so $p-r<p$ and the induction hypothesis applies. For EMVD, the residual problem is a prescribed-norm instance, for which the guarantee has already been established, including when $r=0$.

After reconstructing the ambient coordinates using \eqref{eq:ambient-reconstruction}, the resulting configuration $X_S$ satisfies
\begin{align*}
 \err(X_S)&\le c_S+(1+\eps/4)k_S\\
 &\le(1+\eps/4)(c_S+k_S)\\
 &\le(1+\eps/4)^2k\\
 &\le(1+\eps)k,
\end{align*}
where the last inequality uses $0<\eps\le1$.

Taking the minimum over all candidates and $Y$ proves the approximation guarantee.

\par\smallskip\noindent\textbf{Running time.}\enspace
Each recursive node computes one additive approximation, enumerates a number of subspaces depending only on $p,\eps$, and uses polynomially many compression calls and exact algebraic computations. A path has at most $d+1$ recursive edges: the transition from EMVD to the prescribed-norm problem may leave the dimension unchanged once, but all subsequent calls decrease it. Hence the total number of nodes depends only on $d,\eps$.

For input values of individual encoding length at most $B$ and bounded algebraic degree, \cref{lem:detadd} bounds the output degrees in terms of the parameters and the individual encoding lengths by $f(p,\eps)(B+1+\log(n+1))$, including the input degree bound in $f$. Computing frames, membership tests, votes, residual constraints, and completed coordinates uses only a bounded number of algebraic values per operation. The bounds in \cref{sec:prelim} thus propagate down the recursion and back through the reconstruction of the output. All degrees depend only on $d,\eps$, and every individual encoding length is at most $f(d,\eps)(L+1+\log(n+1))$, for $L$ the original input length. Exact cost evaluation compares one pair at a time and sums the resulting binary losses. The absolute polynomial exponents in \cref{lem:detcompress,lem:detadd} and the bound on the number of nodes give total time $f(d,\eps)(n+L)^{O(1)}$.
\end{proof}

%% file: section3.tex
\section{Hardness in unbounded dimensions}\label{sec:hard}

We reduce from the problem of minimizing the number of unsatisfied rational linear equations~\cite{abss,amaldi}. Given a system of $m$ equations $a_i^\top x=b_i$, where $a_i\in\mathbb Q^p$ and $b_i\in\mathbb Q$ for $i\in[m]$, let $u^*$ denote the minimum number of unsatisfied equations over all assignments $x\in\R^p$.

The idea is to encode the equations using squared distances in $\R^{p+1}$. We introduce reference points at the origin and at $(a_i,0)$ for each equation $i$, together with an additional point $z$, whose position encodes an assignment. If $z=(x,t)$ and $\|z\|^2=R$ for a prescribed positive integer $R$, then
\[
 \|z-(a_i,0)\|^2=R+\|a_i\|^2-2a_i^\top x.
\]
Consequently, subject to $\|z\|^2=R$, the squared distance constraint $\|z-(a_i,0)\|^2=D_i$, where $D_i=R+\|a_i\|^2-2b_i$, holds if and only if $a_i^\top x=b_i$. By choosing $R$ sufficiently large and setting $t=\sqrt{R-\|x\|^2}$, we can impose the norm constraint while retaining an assignment that attains $u^*$.

\begin{figure}[!htbp]
\centering
\begin{tikzpicture}[emvdfig]
\draw[emvdGray!65,line width=.45pt] (4.98,.17)--(4.98,-4.43);
\draw[emvdGray!65,line width=.45pt] (10.08,.17)--(10.08,-4.43);
\begin{scope}
  \node[emvdhead] at (.10,0) {(a) Encode one equation};
  \draw[emvdBlue,line width=.85pt] (.10,-.24)--(4.77,-.24);
  \filldraw[fill=emvdTeal!6,draw=emvdTeal!70,line width=.65pt]
    (.26,-2.12)--(3.57,-2.12)--(4.56,-1.56)--(1.25,-1.56)--cycle;
  \node[emvdnote,font=\scriptsize,text=emvdTeal] at (2.38,-2.30) {$\R^p\times\{0\}$};
  \node[emvdsample] (o) at (.76,-2.04) {};
  \node[emvdsample] (ai) at (3.96,-1.76) {};
  \node[emvderror] (zz) at (2.26,-.86) {};
  \draw[emvdBlue,line width=.75pt] (o)--(zz)--(ai);
  \draw[emvdguide] (2.26,-.86)--(2.26,-1.65);
  \node[emvdnote,anchor=south] at (2.26,-.74) {$z=(x,t)$};
  \node[emvdnote,anchor=east] at (1.36,-1.33) {$R$};
  \node[emvdnote,anchor=west] at (3.17,-1.32) {$D_i$};
  \node[emvdnote,anchor=east] at (.72,-2.05) {$0$};
  \node[emvdnote,anchor=west,font=\scriptsize] at (2.31,-1.61) {$(x,0)$};
  \node[emvdnote,anchor=north west,font=\scriptsize] at (3.78,-1.80) {$(a_i,0)$};
  \node[emvdnote,text width=4.7cm,anchor=north] at (2.43,-2.56)
    {$t=\sqrt{R-\|x\|^2}$\\$D_i=R+\|a_i\|^2-2b_i$\\Both constraints hold iff\\$\|z\|^2=R$ and $a_i^\top x=b_i$.};
\end{scope}
\begin{scope}[shift={(5.10,0)}]
  \node[emvdhead] at (.10,0) {(b) Reference replication};
  \draw[emvdBlue,line width=.85pt] (.10,-.24)--(4.77,-.24);
  \node[emvderror] at (2.43,-.63) {};
  \node[emvdnote,anchor=south] at (2.43,-.54) {$z$};
  \foreach \x/\lab in {.65/0,1.98/1,4.14/m}{
    \draw[emvdBlue!35,fill=emvdLight,rounded corners=.6mm,line width=.55pt]
      (\x-.42,-1.26) rectangle (\x+.42,-2.14);
    \foreach \dx/\dy in {-.22/-1.46,.0/-1.46,.22/-1.46,-.22/-1.69,0/-1.69,.22/-1.69,-.22/-1.92,0/-1.92,.22/-1.92}
      \fill[emvdBlue!50] (\x+\dx,\dy) circle (1.15pt);
    \node[emvdsample] at (\x,-1.46) {};
    \draw[emvdguide] (2.43,-.69)--(\x,-1.39);
    \node[emvdnote] at (\x,-2.37) {$G_{\lab}$};
  }
  \node[emvdnote] at (3.03,-1.69) {$\cdots$};
  \node[emvdnote,font=\scriptsize] at (.65,-2.65) {$m^2$};
  \node[emvdnote,font=\scriptsize] at (1.98,-2.65) {$m$};
  \node[emvdnote,font=\scriptsize] at (4.14,-2.65) {$m$};
  \node[emvdnote,text width=4.65cm,anchor=north] at (2.43,-3.03)
    {Choose one representative\\independently and uniformly\\from each group.};
\end{scope}
\begin{scope}[shift={(10.20,0)}]
  \node[emvdhead] at (.10,0) {(c) Lower bound via sampling};
  \draw[emvdBlue,line width=.85pt] (.10,-.24)--(4.77,-.24);
  \node[emvdnote,text width=4.65cm] at (2.43,-.91)
    {$\displaystyle \E Y=\frac{E_1}{m}$};
  \node[emvdnote,text width=4.65cm] at (2.43,-1.67)
    {$\displaystyle \Prb(\text{bad})\le\frac{E_0+E_2}{m^2}$};
  \draw[emvdarrow] (2.43,-2.06)--(2.43,-2.40);
  \node[emvdnote,text width=4.65cm] at (2.43,-2.70)
    {$u^*\le\E Y+m\Prb(\text{bad})$};
  \node[emvdnote,text width=4.65cm,text=emvdTeal] at (2.43,-3.33)
    {$\displaystyle m u^*\le E_0+E_1+E_2$};
\end{scope}
\node[emvdresult] at (7.55,-4.97)
  {The two directions of the reduction give $\OPT_{p+1}(A)=m u^*$.};
\end{tikzpicture}
\caption{Reduction from $m$ linear equations in $p$ variables.
(a) The reference points lie in $\R^p\times\{0\}$, shown schematically.
The observed squared distances $R$ and $D_i$ encode equation $i$ through the additional point $z$.
(b) The origin group $G_0$ has $m^2$ points, and each equation group $G_i$ has $m$ points. Points within a group coincide in the reference configuration and are drawn separately for clarity.
(c) For an arbitrary configuration, $E_0$ counts disagreements between reference points, $E_1$ those between $z$ and equation copies, and $E_2$ those between $z$ and origin copies. The random variable $Y$ counts disagreements between $z$ and the selected equation representatives. A selection is bad if any observation between selected reference points, or between the selected origin and $z$, is incorrect. Here $u^*$ is the minimum number of unsatisfied equations.}
\label{fig:hardness-roadmap}
\end{figure}
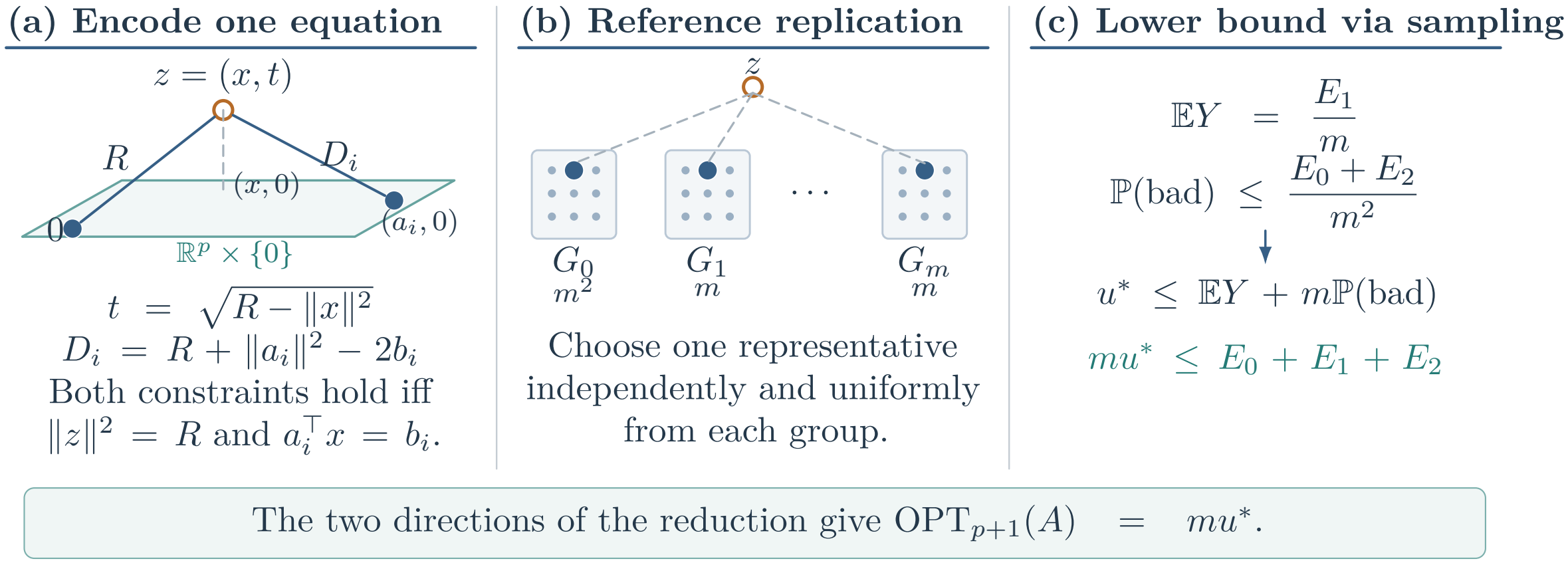

An EMVD solution may change distances not only to $z$, but also between the reference points. To account for these changes, we use a group $G_i$ of $m$ copies of each equation reference point and a group $G_0$ of $m^2$ copies of the origin. We show that the resulting instance $A$ satisfies $\OPT_{p+1}(A)=mu^*$. All observed squared distances between points other than $z$ are jointly realizable in $\R^{p+1}$. \Cref{fig:hardness-roadmap} illustrates the construction and the lower bound argument.

\subsection{The reduction}

\begin{theorem}[Reduction from linear equations]\label{thm:reduction}
Let $a_i^\top x=b_i$, $i\in[m]$, be a system of $m\ge1$ linear equations with $a_i\in\mathbb Q^p$, $b_i\in\mathbb Q$, and $p\ge1$. Let
\[
 u^*=\min_{x\in\R^p}\sum_{i=1}^m\ind[a_i^\top x\ne b_i].
\]
In time polynomial in the encoding length of the system, one can construct an EMVD instance $A$ on $n=2m^2+1$ points in dimension $p+1$ satisfying
\[
 \OPT_{p+1}(A)=mu^*.
\]
All observed squared distances are nonnegative integers. Moreover, there is a designated point $z$ such that all observed squared distances between points other than $z$ are jointly realizable in $\R^{p+1}$, and $\OPT_{p+1}(A)\le m^2=(n-1)/2$.
\end{theorem}

\begin{proof}
\textbf{Construction.}
Clear denominators separately in each equation by multiplying by the least common multiple of its coefficient denominators (including the right-hand side). We continue to write the resulting equations as $a_i^\top x=b_i$, where $a_i\in\mathbb Z^p$ and $b_i\in\mathbb Z$. Let $H\ge1$ be the maximum of one and the absolute values of all resulting coefficients and right-hand sides.

We first show that every consistent subsystem has a solution $x$ satisfying
\[
 \|x\|^2\le p(p!H^p)^2.
\]
Consider a consistent subsystem whose coefficient matrix has rank $r$. If $r=0$, the zero assignment suffices. Otherwise, select $r$ independent coefficient rows and an invertible $r\times r$ submatrix of these rows. Set the variables outside the selected columns to zero and solve the resulting square system. By Cramer's rule, each of the remaining coordinates is a ratio of two integer determinants. The denominator has absolute value at least one, and the numerator has absolute value at most $r!H^r\le p!H^p$. Because the subsystem is consistent, the selected independent equations imply its remaining equations. The resulting assignment therefore satisfies the entire subsystem and has squared norm at most $p(p!H^p)^2$.

Set
\[
 R=1+\max\{p(p!H^p)^2,\,2H\}.
\]
The bit length of $R$ is $O(p(\log(p+1)+\log H))$, which is polynomial in the encoding length of the original system. Clearing denominators also increases the encoding length by at most a polynomial factor.

Let $G_0,G_1,\ldots,G_m$ be disjoint groups of point indices, with $|G_0|=m^2$ and $|G_i|=m$ for $i\in[m]$. In the reference configuration, the points indexed by $G_0$ are placed at the origin and those indexed by $G_i$ at $(a_i,0)\in\R^{p+1}$. Set the observed squared distances between all reference points to their squared Euclidean distances in this configuration. In particular, every observation within a group is zero.

Introduce the additional point $z$. Its observed squared distance to every point in $G_0$ is $R$, and to every point in $G_i$ it is
\begin{equation}\label{eq:linenc}
 D_i=R+\|a_i\|^2-2b_i.
\end{equation}
Since $R>2H$ and $|b_i|\le H$, every observation involving $z$ is positive. The constructed instance has
\[
 n=|G_0|+\sum_{i=1}^m|G_i|+1=2m^2+1
\]
points. All its observations are nonnegative integers of polynomial bit length. The instance can therefore be constructed in polynomial time, and deleting $z$ leaves an exactly realizable squared distance matrix.

\par\smallskip\noindent\textbf{Upper bound.}\enspace
Choose a consistent subsystem of maximum cardinality, which contains $m-u^*$ equations. By the norm bound, it has a solution $x^*$ satisfying $\|x^*\|^2<R$. The maximal cardinality implies that $x^*$ violates exactly $u^*$ equations. Keep all reference points at their positions in the reference configuration and place the additional point at
\[
 (x^*,\sqrt{R-\|x^*\|^2}).
\]
Its squared norm is $R$, so every observation to $G_0$ is satisfied. Its squared distance to a point in $G_i$ is
\[
 R+\|a_i\|^2-2a_i^\top x^*.
\]
By \eqref{eq:linenc}, all $m$ observations to $G_i$ are satisfied if and only if the corresponding equation is satisfied. Since all reference observations are satisfied, this configuration has cost exactly $mu^*$. Thus $\OPT_{p+1}(A)\le mu^*$.

\par\smallskip\noindent\textbf{Lower bound.}\enspace
Consider an arbitrary configuration $X$ in $\R^{p+1}$. Let $E_0$ count disagreements between reference points, $E_1$ disagreements between $z$ and points in $\bigcup_{i=1}^m G_i$, and $E_2$ disagreements between $z$ and points in $G_0$. These classes partition the disagreements, so
\[
 \err_A(X)=E_0+E_1+E_2.
\]

Choose one representative independently and uniformly from each group $G_0,G_1,\ldots,G_m$. Let $q_0,q_1,\ldots,q_m$ denote the positions of these representatives in $X$, and let $q_z$ be the position of $z$. Let $Y$ count disagreements between $q_z$ and the selected equation representatives. Call the selection good if all observations between the selected reference points are satisfied and the observation between $q_0$ and $q_z$ is also satisfied. Otherwise call it bad.

For a good selection, the mutual squared distances of the selected references agree with those of $0,(a_1,0),\ldots,(a_m,0)$. The reference configurations are congruent: their anchored Gram matrices agree, and the isometry between their spans extends to $\R^{p+1}$. Apply the corresponding Euclidean isometry to $q_z$ and denote its new coordinates by $(x,t)$. The observation to the origin gives $\|x\|^2+t^2=R$. For every selected equation representative whose observation to $q_z$ is satisfied, \eqref{eq:linenc} yields
\[
 R+\|a_i\|^2-2a_i^\top x=R+\|a_i\|^2-2b_i.
\]
Thus the assignment $x$ satisfies equation $i$. In particular, $u^*\le Y$ on a good selection. Since $u^*\le m$, for every selection we have
\[
 u^*\le Y+m\ind[\text{selection is bad}].
\]

Each point in an equation group is selected with probability $1/m$, so $\E[Y]=E_1/m$. An incorrect pair between distinct equation groups is selected with probability $1/m^2$, and one between the origin group and an equation group with probability $1/m^3$. Pairs within the same group are never selected together. Each incorrect observation from $z$ to an origin copy is selected with probability $1/m^2$. Hence the union bound gives
\[
 \Prb(\text{selection is bad})\le\frac{E_0+E_2}{m^2}.
\]
Taking expectations, we obtain
\begin{align*}
 u^*&\le \E[Y]+m\Prb(\text{selection is bad})\\
 &\le\frac{E_1}{m}+m\frac{E_0+E_2}{m^2}
 =\frac{\err_A(X)}{m}.
\end{align*}
Since $X$ was arbitrary, $\OPT_{p+1}(A)\ge mu^*$. Together with the upper bound, this proves the cost identity. The remaining properties follow from the construction and $u^*\le m$.
\end{proof}

\subsection{Transferring approximation hardness}

The identity in \cref{thm:reduction} transfers hardness for approximating the optimal value. To handle algorithms that return configurations, we extract an assignment using rational estimates of squared distances and a bound on integer certificates of inconsistency.

\begin{lemma}[Extracting an assignment]\label{lem:hard-extract}
Given the original system of linear equations, the instance $A$ constructed in \cref{thm:reduction}, and an arbitrary configuration $X$ in $\R^{p+1}$ whose coordinates are represented by univariate integer polynomials and rational isolating intervals, one can find a rational assignment violating at most $\err_A(X)/m$ equations. The running time is polynomial in the encoding lengths of the input and $X$.
\end{lemma}

\begin{proof}
Use the integer equations and the parameters $H$ and $R$ from the construction in \cref{thm:reduction}.

\par\smallskip\noindent\textbf{Certificates of inconsistency.}\enspace
Set $K=(p+1)p!H^p$. If a subsystem indexed by $J\subseteq[m]$ is inconsistent, then there exist integers $(c_i)_{i\in J}$ satisfying
\[
 \sum_{i\in J}c_i a_i=0,\qquad
 \sum_{i\in J}c_i b_i\ne0,\qquad
 \sum_{i\in J}|c_i|\le K.
\]
To see this, choose a maximal independent set of coefficient rows from the subsystem and let $r$ be its size. If $r=0$, some equation has a nonzero right-hand side, and a certificate supported on that equation suffices. Otherwise, inconsistency implies that some remaining coefficient row is a linear combination of the selected rows but its right-hand side is not given by the same linear combination. Choose an invertible $r\times r$ submatrix of the selected rows. By Cramer's rule, clearing the common determinant denominator in this linear combination gives an integer dependence on at most $r+1$ rows, with each coefficient of absolute value at most $r!H^r\le p!H^p$. The corresponding combination of right-hand sides is nonzero. Since $r\le p$, the sum of the absolute values is at most $K$, proving the claim.

\par\smallskip\noindent\textbf{Identifying disagreements.}\enspace
Set
\[
 \rho=\frac{1}{64K^2(R+1)}.
\]
For every unordered pair $\{i,j\}$ of points in $X=(x_i)_{i\in V}$, compute a rational estimate $\widehat D_{ij}$ satisfying
\[
 |\widehat D_{ij}-\|x_i-x_j\|^2|\le\rho/2.
\]
Flag the pair if $|\widehat D_{ij}-A_{ij}|>\rho/2$. Every flagged pair is a disagreement, since an exactly satisfied observation cannot pass this threshold. Every unflagged pair satisfies
\[
 |\|x_i-x_j\|^2-A_{ij}|\le\rho.
\]

These estimates can be computed in polynomial time in the input and output encoding lengths. Indeed, the bit lengths of $K$, $R$, and $1/\rho$ are polynomial in the input length. Standard root bounds for the defining univariate polynomials give an upper bound $M\ge1$ on the absolute values of all coordinates of $X$, whose logarithm is polynomial in the output length. For example, approximating each coordinate to within $\rho/(32(p+1)(M+1))$ suffices to compute every squared distance to absolute accuracy $\rho/2$, by expanding the squares. Univariate root isolation computes such rational coordinate approximations with polynomial bit complexity.

\par\smallskip\noindent\textbf{Selecting representatives.}\enspace
As in the reduction, independently and uniformly select one representative from each group $G_0,G_1,\ldots,G_m$. Let $Y'$ count flagged pairs from $z$ to selected equation representatives, and let $N'$ count flagged pairs between selected references together with the indicator that the selected origin has a flagged observation to $z$. By the selection probabilities established in the proof of \cref{thm:reduction}, and since every flagged pair is a disagreement,
\[
 \E[Y'+mN']\le\frac{\err_A(X)}{m}.
\]
Each term of $Y'+mN'$ depends on at most two group choices; its conditional expectation can be computed exactly by rational arithmetic. Fix one representative at a time without increasing the conditional expectation. The resulting selection satisfies
\[
 Y'+mN'\le\frac{\err_A(X)}{m}.
\]
If $N'>0$, then the left-hand side is at least $m$, so any rational assignment satisfies the claimed guarantee. It remains to consider $N'=0$.

\par\smallskip\noindent\textbf{Consistency of the selected equations.}\enspace
Let $q_0,q_1,\ldots,q_m$ and $q_z$ be the positions of the selected references and the additional point in $X$. Let $J\subseteq[m]$ consist of the equation indices for which the pair between the selected representative and $z$ is unflagged. Then $|J|=m-Y'$. For $i\in J$, set $v_i=q_i-q_0$ and $w=q_z-q_0$. Since $N'=0$, every observed squared distance between selected references differs from its actual value by at most $\rho$, and $|\|w\|^2-R|\le\rho$. By the polarization identity,
\[
 |\langle v_i,v_j\rangle-a_i^\top a_j|\le 3\rho/2,
 \qquad i,j\in J.
\]
Also, the squared distance between $q_i$ and $q_z$ differs from $D_i$ by at most $\rho$ for $i\in J$. Using \eqref{eq:linenc} in
\[
 \langle v_i,w\rangle=\tfrac12\bigl(\|q_i-q_0\|^2+\|q_z-q_0\|^2-\|q_i-q_z\|^2\bigr)
\]
gives
\[
 |\langle v_i,w\rangle-b_i|\le 3\rho/2,\qquad
 \|w\|^2\le R+\rho.
\]

Suppose the subsystem indexed by $J$ is inconsistent. The certificate above gives integers $(c_i)_{i\in J}$ with $\sum_i c_i a_i=0$, $\sum_i|c_i|\le K$, and $|\sum_i c_i b_i|\ge1$. Using the estimates of the reference inner products,
\begin{align*}
 \left\|\sum_{i\in J}c_i v_i\right\|^2
 &=\sum_{i,j\in J}c_i c_j\bigl(\langle v_i,v_j\rangle-a_i^\top a_j\bigr)\\
 &\le\tfrac32\rho\left(\sum_{i\in J}|c_i|\right)^2
 \le\tfrac32\rho K^2.
\end{align*}
Combining this estimate with the bounds on $\langle v_i,w\rangle$ and applying the Cauchy--Schwarz inequality gives
\begin{align*}
 1\le\left|\sum_{i\in J}c_i b_i\right|
 &\le \left|\left\langle\sum_{i\in J}c_i v_i,w\right\rangle\right|+\tfrac32\rho K\\
 &\le K\sqrt{\tfrac32\rho(R+\rho)}+\tfrac32\rho K\\
 &\le\sqrt{3/128}+3/128<1,
\end{align*}
where the last line uses the definition of $\rho$, $K\ge1$, and $\rho\le1$. Thus the subsystem indexed by $J$ is consistent.

\par\smallskip\noindent\textbf{Extracting an assignment.}\enspace
Solve this subsystem by rational linear algebra. The resulting assignment violates at most $m-|J|=Y'$ equations, and
\[
 Y'\le Y'+mN'\le\frac{\err_A(X)}{m}.
\]
All steps take polynomial time in the input and output encoding lengths.
\end{proof}

\begin{proof}[Proof of \cref{thm:hard}]
Arora et al.~\cite{abss} (Theorems 5 and 7; see also Amaldi and Kann~\cite{amaldi}) show that minimizing the number of unsatisfied rational linear equations is NP-hard to approximate within any fixed constant factor greater than one. Moreover, for every fixed $0<\beta<1$, a polynomial-time approximation within a factor of $2^{(\log N_0)^{1-\beta}}$, where $N_0$ is the encoding length of the system, would imply that every language in $\mathsf{NP}$ has a deterministic quasipolynomial-time decision algorithm. The hardness applies to rational assignments; allowing real assignments does not change the optimum, because a consistent rational linear system has a rational solution.

Suppose a deterministic polynomial-time EMVD algorithm returns a configuration $X$ with
\[
 \err_A(X)\le\rho(n)\OPT_{p+1}(A)
\]
on instances of the reduction. By \cref{thm:reduction,lem:hard-extract}, we can extract a rational assignment in polynomial time that violates at most
\[
 \frac{\err_A(X)}{m}\le\rho(n)u^*
\]
equations. Thus the approximation ratio is preserved, giving the constant-factor hardness.

For the stronger bound, fix $0<\alpha<1$ and suppose that $\rho(n)=2^{(\log n)^{1-\alpha}}$. Since $m\le N_0$ and $n=2m^2+1$, we have $\log n=O(\log N_0)$. Consequently, for sufficiently large $N_0$,
\[
 (\log n)^{1-\alpha}\le(\log N_0)^{1-\alpha/2}.
\]
The resulting algorithm for the linear equation problem has approximation ratio at most $2^{(\log N_0)^{1-\alpha/2}}$. The finitely many smaller input sizes can be handled by exact optimization. Since $0<\alpha/2<1$, the result of Arora et al.~\cite{abss} implies that $\mathsf{NP}$ has deterministic quasipolynomial-time algorithms.

By \cref{thm:reduction}, both hardness bounds hold on instances with a single exceptional point and $\OPT_{p+1}(A)\le(n-1)/2$.
\end{proof}\textbf{}

%% file: section4.tex
\section{Approximation in arbitrary dimensions}
\label{sec:pivots}

We now give a randomized approximation algorithm with polynomial running time in both the input length and dimension. Its guarantee depends on the affine dimension $r\le d$ of an arbitrary comparison configuration $X^*=(x_i^*)_{i\in V}$, of cost $k=\err_A(X^*)$. The algorithm does not need $X^*$, $r$, or $k$.

\paragraph{High-level idea.}
Starting with a uniformly chosen anchor at the origin, we construct one coordinate direction at a time. Each pivot is chosen uniformly from points whose positions have not yet been determined. Its observed distances give proposed coordinates for the other unresolved points. If a proposal is infeasible, we place the point within the enlarged span. All updates preserve distances to previously resolved points, allowing us to bound the propagation of errors. The single-run analysis gives an expected cost of $O((r+1)^2)k$. We obtain a better guarantee by independent runs, as described in \cref{sec:pivot-amplify}.

\subsection{Random pivot reconstruction}\label{sec:pivot-algorithm}

Let $D=\min\{d,n-1\}$. Since any configuration of $n$ points has affine dimension at most $n-1$, at most $D$ coordinate directions are needed. If $n=1$, we return the origin.

\paragraph{Initialization.}
Choose an initial anchor $a_0$ uniformly from $[n]$ and place it at the origin. For every $i\in V$, set
\[
 u_i=(),\qquad R_i=A_{a_0i},
\]
where $A_{a_0a_0}=0$. At the beginning of iteration $t$, let $S_{t-1}$ be the subspace spanned by the first $t-1$ coordinate directions. The vector $u_i\in\R^{t-1}$ contains the coordinates of point $i$ that have already been determined, and $R_i\ge0$ is the squared norm to be assigned to its remaining coordinates. Equivalently, $R_i$ is the squared distance of any completion of point $i$ to $S_{t-1}$. We call a point \emph{resolved} if $R_i=0$, since its position is then completely determined. Let $U=\{i\in V:R_i>0\}$ be the set of active points. Resolved points receive zero in every subsequent coordinate.

\paragraph{Choosing a pivot and extending the coordinates.}
If $U$ is empty, the reconstruction is complete. Otherwise, choose a pivot $a$ uniformly from $U$. We place its residual vector along the new coordinate direction, giving it coordinate $\sqrt{R_a}$. For an active point $i$, suppose its coordinate along this direction is $z_i$. If the coordinate is feasible, its remaining squared norm is $R_i-z_i^2$, and its squared distance to the pivot is
\[
 \|u_i-u_a\|^2+(z_i-\sqrt{R_a})^2+(R_i-z_i^2)
 =\|u_i-u_a\|^2+R_a+R_i-2\sqrt{R_a}z_i.
\]
Equating this expression with the observed squared distance $A_{ai}$ gives the proposed coordinate
\begin{equation}\label{eq:pivot}
 H_{ai}=\frac{R_a+R_i+\|u_a-u_i\|^2-A_{ai}}{2},\qquad
 z_i=\frac{H_{ai}}{\sqrt{R_a}}.
\end{equation}
We compute these proposals for all active points before updating any of their positions or residual norms.

\begin{figure}[!htbp]
\centering
\begin{tikzpicture}[emvdfig]
\draw[emvdGray!65,line width=.45pt] (4.98,.17)--(4.98,-4.02);
\draw[emvdGray!65,line width=.45pt] (10.08,.17)--(10.08,-4.02);
\begin{scope}
  \node[emvdhead] at (.10,0) {(a) Choose a pivot};
  \draw[emvdBlue,line width=.85pt] (.10,-.24)--(4.77,-.24);
  \draw[emvdarrow] (.55,-2.11)--(4.50,-2.11) node[anchor=west,inner sep=1pt] {$e_t$};
  \draw[emvdBlue,line width=.8pt,-{Latex[length=1.8mm]}] (.76,-2.11)--(3.93,-2.11);
  \draw[emvdTeal,line width=.9pt,-{Latex[length=1.8mm]}] (.76,-2.11)--(2.51,-.79);
  \draw[emvdguide] (2.51,-.79)--(2.51,-2.11);
  \node[emvdsample] at (3.93,-2.11) {};
  \node[emvdpoint] at (2.51,-.79) {};
  \fill[emvdInk] (.76,-2.11) circle (1.25pt);
  \node[emvdnote,anchor=south] at (2.51,-.72) {$w_i$};
  \node[emvdnote,anchor=south] at (3.93,-2.04) {$w_a$};
  \node[emvdnote] at (2.51,-2.37) {$z_i$};
  \node[emvdnote,text width=4.6cm,anchor=north] at (2.43,-2.66)
    {Consistent case shown.\\$\displaystyle z_i=H_{ai}/\sqrt{R_a}$};
\end{scope}
\begin{scope}[shift={(5.10,0)}]
  \node[emvdhead] at (.10,0) {(b) Update or freeze};
  \draw[emvdBlue,line width=.85pt] (.10,-.24)--(4.77,-.24);
  \draw[emvdGray,line width=.65pt] (.25,-1.15)--(4.60,-1.15);
  \draw[emvdTeal,line width=2pt] (.88,-1.15)--(3.79,-1.15);
  \foreach \x in {.88,3.79}{\draw[emvdTeal,line width=.7pt] (\x,-1.26)--(\x,-1.04);}
  \node[emvdpoint] at (2.40,-1.15) {};
  \node[emvderror] at (4.43,-1.15) {};
  \node[emvderror] at (.30,-1.15) {};
  \draw[emvdAmber,line width=.7pt,-{Latex[length=1.5mm]}]
    (.30,-.94) to[bend left=28] (3.79,-.94);
  \node[emvdnote,anchor=south] at (2.40,-1.08) {$z_i$};
  \node[emvdnote,anchor=south,text=emvdAmber,font=\scriptsize] at (4.18,-.65) {infeasible};
  \draw[emvdAmber,line width=.7pt,-{Latex[length=1.5mm]}] (4.42,-.94) to[bend right=30] (3.79,-.94);
  \node[emvdnote,anchor=north] at (.88,-1.31) {$-\sqrt{R_i}$};
  \node[emvdnote,anchor=north] at (3.79,-1.31) {$\sqrt{R_i}$};
  \node[emvdnote,text width=4.65cm,font=\scriptsize,anchor=north] at (2.43,-1.82)
    {$c_i=z_i$ if $z_i^2\le R_i$.\\If $t=D$, require equality.};
  \node[emvdnote,text width=4.65cm,font=\scriptsize,text=emvdAmber!90!black,anchor=north] at (2.43,-2.72)
    {Otherwise $c_i=+\sqrt{R_i}$.\\Always $R_i\gets R_i-c_i^2$.};
\end{scope}
\begin{scope}[shift={(10.20,0)}]
  \node[emvdhead] at (.10,0) {(c) Fixed distances};
  \draw[emvdBlue,line width=.85pt] (.10,-.24)--(4.77,-.24);
  \draw[emvdTeal,line width=.95pt] (.33,-1.78)--(4.46,-1.78);
  \node[emvdsample] (j) at (.80,-1.78) {};
  \node[emvdpoint] (i1) at (3.30,-.81) {};
  \node[emvdpoint] (i2) at (3.30,-2.75) {};
  \draw[emvdguide] (3.30,-.81)--(3.30,-2.75);
  \draw[emvdBlue,line width=.7pt] (j)--(i1) (j)--(i2);
  \node[emvdnote,anchor=north west] at (.80,-1.84) {$x_j$};
  \node[emvdnote,anchor=west] at (3.36,-.81) {$x_i$};
  \node[emvdnote,anchor=west] at (3.36,-2.75) {$x_i'$};
  \node[emvdnote,anchor=east] at (3.25,-1.54) {$u_i$};
  \node[emvdnote,font=\scriptsize,text=emvdTeal] at (1.18,-.84) {earlier span};
  \draw[emvdguide] (.67,-.99)--(.48,-1.75);
  \node[emvdnote,anchor=west] at (3.36,-1.32) {$\sqrt{R_i}$};
  \node[emvdnote,text width=4.6cm,anchor=north] at (2.43,-3.03)
    {Both squared distances equal\\$\|u_i-u_j\|^2+R_i$.};
\end{scope}
\node[emvdresult] at (7.55,-4.55)
  {Repeat until all points are resolved or $D$ axes are used; keep the best of independent runs.};
\end{tikzpicture}
\caption{Random pivot reconstruction (schematic). (a) When the observation is consistent, the proposed coordinate $z_i$ equals the projection of the residual vector $w_i$ onto the new direction $e_t$, aligned with the pivot residual $w_a$. (b) If the proposed coordinate lies outside $[-\sqrt{R_i},\sqrt{R_i}]$, the algorithm uses the positive endpoint $+\sqrt{R_i}$, regardless of the sign of the proposal. On the final direction $t=D$, the ordinary update is accepted only when $z_i^2=R_i$. The actual coordinate $c_i$ follows \eqref{eq:pivot-update}, and the squared residual norm becomes $R_i-c_i^2$. (c) Two possible completions $x_i$ and $x_i'$ with the same earlier projection $u_i$ and squared residual norm $R_i$ have the same squared distance to a previously resolved point $x_j$. This illustrates the invariant preserved by every update.}
\label{fig:pivot-roadmap}
\end{figure}
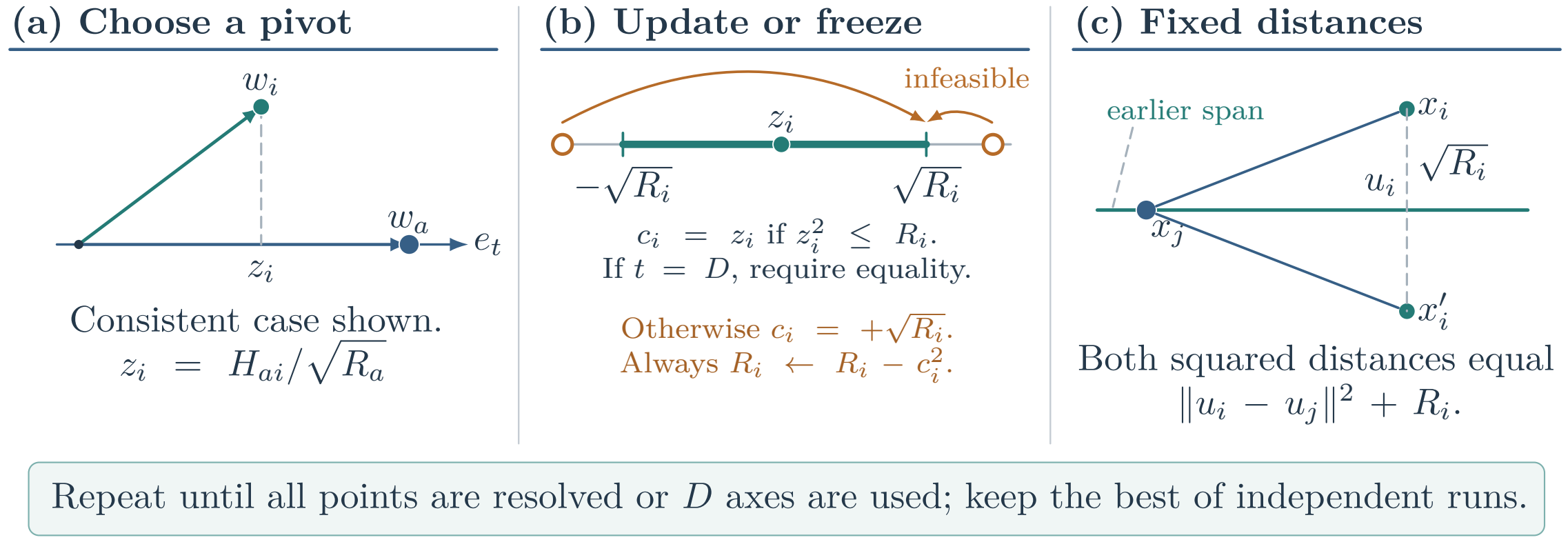

\paragraph{Updating and freezing.}
If $z_i^2\le R_i$, the proposed coordinate is feasible, provided that there is a remaining coordinate direction or that $z_i^2=R_i$. Otherwise, we use the positive coordinate $+\sqrt{R_i}$, which places the point in the enlarged subspace $S_t$. We call this operation \emph{freezing}. In particular, on the final direction $t=D$ we accept the ordinary update only when $z_i^2=R_i$. For an active point $i$, the actual coordinate $c_i$ is therefore
\begin{equation}\label{eq:pivot-update}
 c_i=\begin{cases}
 z_i, & t<D\text{ and }z_i^2\le R_i,\\
 z_i, & t=D\text{ and }z_i^2=R_i,\\
 +\sqrt{R_i}, & \text{otherwise}.
 \end{cases}
\end{equation}
For a resolved point, set $c_i=0$. In both cases, append $c_i$ to $u_i$ and set $R_i\gets R_i-c_i^2$. A frozen point has zero residual norm and receives zero in all subsequent coordinates. For the pivot itself, $A_{aa}=0$ and \eqref{eq:pivot} gives $z_a=\sqrt{R_a}$, so the pivot becomes resolved through an ordinary update. \Cref{alg:pivots} gives the reconstruction procedure, and \cref{fig:pivot-roadmap} illustrates the update.

\begin{algorithm}[!ht]
\caption{One random pivot reconstruction}\label{alg:pivots}
\small
\makeatletter
\def\theHALG@line{\thealgorithm.\arabic{ALG@line}}
\makeatother
\begin{algorithmic}[1]
\Require Nonnegative rational squared distance observations $A$ on $n\ge1$ points, dimension $d\ge1$
\Ensure Configuration $X\in(\R^d)^n$
\State $D\gets\min\{d,n-1\}$; choose $a_0$ uniformly from $[n]$.
\State Set $u_i\gets()$ and $R_i\gets A_{a_0i}$ for every $i\in[n]$.
\For{$t=1,\ldots,D$}
  \State $U\gets\{i\in[n]:R_i>0\}$
  \If{$U=\varnothing$}
    \State \textbf{break}
  \EndIf
  \State Choose $a$ uniformly from $U$.
  \State Compute $z_i$ for all $i\in U$ using \eqref{eq:pivot}.
  \ForAll{$i\in[n]$}
    \If{$R_i=0$}
      \State $c_i\gets0$
    \ElsIf{$z_i^2>R_i$ or $(t=D\text{ and }z_i^2<R_i)$}
      \State $c_i\gets+\sqrt{R_i}$ \Comment{Freeze}
    \Else
      \State $c_i\gets z_i$
    \EndIf
    \State Append $c_i$ to $u_i$; set $R_i\gets R_i-c_i^2$.
  \EndFor
\EndFor
\State Pad every $u_i$ with zeros to dimension $d$; return $(u_i)_{i\in[n]}$.
\end{algorithmic}
\end{algorithm}

\paragraph{Geometric invariant.}
Let $u_i'$ and $R_i'$ denote the projection coordinates and squared residual norm of point $i$ after an update. By construction,
\[
 u_i'=(u_i,c_i),\qquad c_i^2+R_i'=R_i.
\]
Thus every update preserves the projection onto the previous subspace and the squared distance to that subspace. Consider a point $j$ that was resolved before the update. Since $R_j=0$, its new coordinate is zero and its position remains unchanged. The squared distance between $i$ and $j$ after the update satisfies
\[
 \|u_i'-u_j'\|^2+R_i'
 =\|u_i-u_j\|^2+c_i^2+R_i'
 =\|u_i-u_j\|^2+R_i.
\]
Consequently, once either endpoint of a pair is resolved, its squared distance to the other endpoint is fixed throughout the remaining reconstruction, whether or not that distance agrees with the observation. Each iteration resolves at least the selected pivot, and the final iteration resolves all remaining active points. The algorithm therefore terminates after at most $D$ iterations and returns a configuration in $\R^d$.

\paragraph{Independent runs.}
The final algorithm performs $M=\lceil4\sqrt{2n}\rceil$ independent runs of \cref{alg:pivots}, evaluates the cost of every resulting configuration on all unordered pairs, and returns one of minimum cost. We establish the approximation guarantee in \cref{sec:pivot-expectation,sec:pivot-amplify}.

\begin{lemma}[Polynomial bit complexity of pivoting]\label{lem:pivotbit}
For rational squared distance observations, the reconstruction in \cref{alg:pivots}, given its sequence of pivot choices, and the exact evaluation of its output cost take $\poly(n,d,L)$ time.
\end{lemma}
\begin{proof}
Let $a_0,a_1,\ldots,a_t$ be the anchors selected so far, including the initial anchor. Each new pivot has positive residual norm when selected and has satisfied all its observations to the previous anchors, since an active point has never frozen. Thus the selected anchors are affinely independent, and all observations between them are satisfied. Let $G_t$ be the anchored Gram matrix of these anchors, defined as in \eqref{eq:gram}. It is rational and positive definite. Adding a new pivot extends the Gram matrix by a positive Schur complement equal to the pivot's squared residual norm.

Write the factorization of $G_t$ as
\[
 G_t=L_G\operatorname{diag}(\Delta_1,\ldots,\Delta_t)L_G^\top,
\]
where $L_G$ is a rational unit lower triangular matrix and each $\Delta_j$ is a positive rational number. These diagonal entries are the squared residual norms of the successive pivots. For an active point $i$, the vector of its inner products with the anchors is rational by \eqref{eq:gram}. Solving the triangular system associated with this factorization gives its coordinate in direction $j$ in the form
\[
 u_{ij}=\beta_{ij}\sqrt{\Delta_j},\qquad \beta_{ij}\in\Q.
\]
Indeed, the triangular system has coefficient matrix $L_G\operatorname{diag}(\sqrt{\Delta_1},\ldots,\sqrt{\Delta_t})$. Its solution has each coordinate equal to a rational number divided by $\sqrt{\Delta_j}$, which has the asserted form because $\Delta_j$ is rational and positive. The squared residual norm of an active point is its initial rational squared norm minus the sum of the squares of these coordinates, so it is rational. Determinant bounds for rational linear systems and their Schur complements give polynomial bit lengths for all these rational quantities.

A point that freezes has coordinates of the above form before its freezing step. At that step its coordinate is the positive square root of its remaining rational squared norm; all later coordinates are zero. A point resolved by an ordinary update has the usual coordinates followed by zeros. We store each coordinate as a rational multiple of a square root of a nonnegative rational number. Such a number has an exact algebraic representation of polynomial bit length, as required by \cref{sec:prelim}. All comparisons and updates of residual norms use rational arithmetic.

Consider the squared distance between any two output points. The squared coordinate terms are rational. Products of ordinary coordinates along the same direction are also rational, since
\[
 (\beta_{ij}\sqrt{\Delta_j})(\beta_{hj}\sqrt{\Delta_j})
 =\beta_{ij}\beta_{hj}\Delta_j.
\]
If either point freezes, at most one cross term can involve an additional square root, namely the term at the earlier freezing step (or the common freezing step if both freeze together). All later cross terms vanish because the frozen point has zero subsequent coordinates. Hence each squared distance is of the form $a+b\sqrt c$ for rationals $a,b,c$ of polynomial bit length with $c\ge0$. Equality to a rational observation $A_{ih}$ can be tested by comparing the signs of $A_{ih}-a$ and $b\sqrt c$ and checking $(A_{ih}-a)^2=b^2c$. The sign of $b\sqrt c$ is determined by $b$ and whether $c=0$. Thus we need only exact rational arithmetic.

There are at most $D\le d$ coordinate updates and $\binom{n}{2}$ pairs to evaluate. Both reconstruction and cost evaluation therefore have polynomial bit complexity.
\end{proof}

\subsection{Expected cost of a single run}\label{sec:pivot-expectation}

For the fixed comparison configuration $X^*$, define the error graph $G^*=(V,E^*)$ by
\[
 E^*=\{\{i,j\}:1\le i<j\le n,\ A_{ij}\ne\|x_i^*-x_j^*\|^2\}.
\]
Thus $|E^*|=k$.

\paragraph{Clean and contaminated points.}
We mark points that may have been affected by erroneous observations. Initially, mark every neighbor of the initial anchor $a_0$ in $G^*$. Whenever an unmarked pivot is selected, mark all its unmarked active neighbors in $G^*$. Once marked, a point remains marked. We call marked points \emph{contaminated} and unmarked points \emph{clean}. A contaminated point may still be placed correctly. The marks are used only in the analysis.

Until the first contaminated pivot is selected, every clean point has the same projection and squared residual norm as its position in $X^*$, in a common coordinate frame. Initially, translate $X^*$ so that $x_{a_0}^*=0$. Every clean point has a correct observation to $a_0$, so its initial squared residual norm is correct. Suppose the invariant holds before a clean pivot is selected. Align the comparison residual of this pivot with the new coordinate direction. For every still clean active point whose observation to the pivot is correct, \eqref{eq:pivot} gives exactly its comparison coordinate. The ordinary update is feasible and preserves the invariant. Clean resolved points receive zero in the new coordinate. If the final direction is reached after only clean pivots, the remaining clean points have zero residual norm after the ordinary update, since $r\le D$. More generally, after $r$ clean pivot directions have been selected, every clean point is resolved because the comparison configuration has affine dimension $r$.

\paragraph{Charging disagreements.}
Let $U_t$ be the active set immediately before pivot step $t\ge1$, and let $m_t=|U_t|$. Let $N_t$ be the set of points newly marked at step $t$ when a clean pivot is selected. Treat the initial marking as step zero, with $m_0=n$ and $N_0$ the set of neighbors of $a_0$ in $G^*$. Define the charge
\[
 C_t=m_t|N_t|.
\]
For $t\ge1$, this bounds the number of pairs whose endpoints were both active at the beginning of step $t$ and at least one of which is newly marked. The initial charge $C_0$ covers all pairs incident to the initially marked points.

Consider a pair outside $E^*$. If both endpoints are clean when one first becomes resolved, their distance agrees with the comparison distance and remains unchanged throughout the rest of the reconstruction. Otherwise, if an endpoint is marked before or during the step at which either endpoint first becomes resolved, the pair is covered by the charge at the first such marking: both endpoints were active just before that step, or the marking occurred at initialization. This accounts for every additional disagreement whose distance is fixed before the first contaminated pivot is selected.

If a contaminated pivot is selected for the first time at step $t$, charge all remaining active pairs, at most $\binom{m_t}{2}$. Since every distance involving a previously resolved point is fixed, no other uncharged disagreement can arise after this step. If the reconstruction reaches $r$ clean pivot directions without selecting a contaminated pivot, every remaining active point is contaminated, so every pair between these points has already been charged. If the active set is empty earlier, all distances have already been fixed. Thus these charges, together with the original $k$ disagreements, bound the cost of the final configuration. The same argument allows us to stop the analysis earlier and charge all remaining active pairs, regardless of subsequent pivot choices.

\paragraph{Expected charges.}
The initial anchor is uniform, so
\[
 \E[C_0]=\frac1n\sum_{a\in V}n\deg_{G^*}(a)=2k.
\]
At a later step $t$, condition on the history before selecting the pivot, assuming the analysis has not stopped. Set $C_t=0$ if the selected pivot is contaminated. Otherwise, the number of newly marked points is at most the degree of the pivot in the induced subgraph $G^*[U_t]$. Consequently,
\begin{equation}\label{eq:expectedC}
 \E[C_t\mid\text{history}]
 \le\frac1{m_t}\sum_{a\in U_t}m_t\deg_{G^*[U_t]}(a)
 \le2k.
\end{equation}
Let $b_t$ denote the number of contaminated points still active before step $t$. Each such point was marked at some earlier step $j<t$. Since the active sets only shrink, $m_j\ge m_t$. Therefore,
\begin{equation}\label{eq:contam}
 b_t m_t\le\sum_{j<t}C_j.
\end{equation}
The conditional probability of selecting a contaminated pivot is $b_t/m_t$. Let $F_t$ be the additional charge at this step if the first contaminated pivot is selected, and zero otherwise. Its conditional expectation is bounded by
\[
 \E[F_t\mid\text{history}]
 \le\frac{b_t}{m_t}\binom{m_t}{2}
 \le\frac{b_t m_t}{2}
 \le\frac12\sum_{j<t}C_j.
\]
Extend all charges by zero after the first contaminated pivot or after the analysis terminates. Taking expectations and using \eqref{eq:expectedC} gives $\E[F_t]\le tk$. There are at most $r$ pivot steps in the analysis, in addition to the initial marking. Combining the charges yields
\begin{align*}
 \E[\err_A(X)]
 &\le k+\sum_{t=0}^r\E[C_t]+\sum_{t=1}^r\E[F_t]\\
 &\le k+2(r+1)k+\frac{r(r+1)}{2}k\\
 &=\frac{(r+2)(r+3)}{2}k.
\end{align*}

\subsection{Improving the approximation by independent runs}\label{sec:pivot-amplify}

To improve the approximation ratio, we compare the actual reconstruction with an auxiliary process that selects only clean pivots.

\begin{proof}[Proof of the approximation guarantee in \cref{thm:dim}]
If $k=0$, every point remains clean and the reconstruction is exact. Assume $k>0$ and set
\[
 q=r+1,\qquad B=16qk,\qquad
 \alpha=64\left(q+\frac{q^2}{\log(2n)}\right).
\]

\paragraph{An auxiliary process.}
Consider a process that chooses the initial anchor uniformly, as in \cref{alg:pivots}, but selects each subsequent pivot uniformly from the clean active points. Use the same marking rules and charges $C_t$ as in the preceding analysis. Stop before the next pivot if any of the following conditions holds:
\begin{enumerate}
 \item The cumulative charge exceeds $B$.
 \item The number of active points satisfies $m_t^2\le\alpha k$.
 \item The process has selected $r$ clean pivot directions.
 \item The active set is empty.
\end{enumerate}
The second condition ensures that all remaining active pairs can be charged at an additional cost of at most $\alpha k/2$. Neither the auxiliary process nor its stopping conditions are used by the algorithm.

Whenever the process has not stopped, \eqref{eq:contam} gives
\[
 \frac{b_t}{m_t}\le\frac{\sum_{j<t}C_j}{m_t^2}
 \le\frac{B}{\alpha k}\le\frac14.
\]
Hence a clean active pivot exists. Conditional on the history, the expected new charge satisfies
\begin{align*}
 \E[C_t\mid\text{history}]
 &\le\frac{m_t}{m_t-b_t}\sum_{a\in U_t}\deg_{G^*[U_t]}(a)\\
 &\le\frac{2k}{1-b_t/m_t}\le\frac{8k}{3}.
\end{align*}
There are at most $r$ pivot steps and the expected initial charge is $2k$. Thus the expected cumulative charge at stopping is at most $2k+8rk/3\le3qk$. Markov's inequality gives probability at least
\[
 1-\frac{3qk}{B}=\frac{13}{16}
\]
that the cumulative charge at stopping is at most $B$. Call a stopped sequence of pivots favorable if it satisfies this bound.

\paragraph{Probability of a favorable sequence.}
Fix a favorable sequence generated by the auxiliary process. At every step after the initial anchor, the probability of selecting its next pivot is $1/(m_t-b_t)$ in the auxiliary process and $1/m_t$ in the actual algorithm. The ratio of these probabilities is $1-b_t/m_t$. Since the initial anchor has the same distribution in both processes, the probability that the actual algorithm follows this sequence as a prefix is its auxiliary probability multiplied by
\[
 \prod_t\left(1-\frac{b_t}{m_t}\right).
\]
Before each selected pivot, $b_t/m_t\le B/(\alpha k)\le1/4$. Using $\log(1-x)\ge-2x$ for $0\le x\le1/2$ and the bound of $r$ on the number of steps, we obtain
\begin{align*}
 \prod_t\left(1-\frac{b_t}{m_t}\right)
 &\ge\exp\left(-2\sum_t\frac{b_t}{m_t}\right)\\
 &\ge\exp\left(-\frac{2rB}{\alpha k}\right)\\
 &\ge\exp\left(-\frac{\log(2n)}{2}\right)
 =\frac{1}{\sqrt{2n}}.
\end{align*}
The last inequality follows from
\[
 \frac{2rB}{\alpha k}
 =\frac{rq}{2(q+q^2/\log(2n))}
 \le\frac{\log(2n)}2,
\]
where $r\le q$. The stopping rule depends only on the pivot sequence and the fixed comparison configuration. Thus distinct stopped sequences define disjoint events when viewed as prefixes of the actual reconstruction. Summing over all favorable sequences, the actual algorithm follows a favorable prefix with probability at least $13/(16\sqrt{2n})$.

\paragraph{Cost after a favorable prefix.}
At a favorable stopping point, the cumulative charge is at most $B$, so the first stopping condition does not hold. If $r$ clean directions have been selected, every remaining active point is contaminated, and all remaining active pairs have already been charged. If the active set is empty, no further disagreements can arise. Otherwise, the process stopped because $m_t^2\le\alpha k$. By the geometric invariant, arbitrary continuation of the reconstruction can add at most $\binom{m_t}{2}\le\alpha k/2$ disagreements beyond the reserved charges. In all cases the final cost is at most
\[
 k+B+\frac{\alpha k}{2}
 =O\left(r+1+\frac{(r+1)^2}{\log(2n)}\right)k.
\]

\paragraph{Independent runs.}
Perform $M=\lceil4\sqrt{2n}\rceil$ independent reconstructions. The probability that none follows a favorable prefix is at most
\[
 \left(1-\frac{13}{16\sqrt{2n}}\right)^M
 \le\exp\left(-\frac{13M}{16\sqrt{2n}}\right)
 \le e^{-13/4}<\frac13.
\]
By \cref{lem:pivotbit}, each reconstruction and its cost evaluation take polynomial time given the pivot choices. To implement the algorithm with fair random bits and worst-case polynomial running time, cap rejection sampling at $\lceil\log_2(12nM)\rceil$ trials for each uniform choice. Return an arbitrary feasible configuration if any choice fails. Each trial succeeds with probability at least $1/2$, and there are at most $nM$ choices, so truncation adds at most $1/12$ to the failure probability. The total failure probability is at most $e^{-13/4}+1/12<1/3$. The expectation and lower bounds refer to exact uniform pivot sampling.

Selecting the configuration of minimum cost among the $M$ outputs therefore gives the claimed approximation guarantee with probability at least $2/3$ in polynomial time. Since $r\le d$ and $d\ge1$, this gives an $O(d+d^2/\log(2n))$-approximation, and an $O(d)$-approximation when $d=O(\log n)$.
\end{proof}

\subsection{A lower bound for random pivot reconstruction}

We show that polynomially many independent runs of \cref{alg:pivots} cannot substantially improve its dependence on the dimension. The construction changes only the distances corresponding to a perfect matching. Nevertheless, whenever the first $d$ anchors contain both endpoints of a matching edge, the reconstruction produces $\Omega(n^2)$ disagreements.

\begin{theorem}[Lower bound for uniform anchor reconstruction]\label{thm:pivotlower}
Fix any constant $C>0$. For arbitrarily large $d$, there are EMVD instances with $n=\Theta_C(d^2/\log d)$ points and positive rational observed squared distances between distinct points, such that the best of at most $n^C$ independent runs of \cref{alg:pivots} has approximation ratio
\[
 \Omega_C\left(\frac{d^2}{\log d}\right)
\]
with probability tending to one as $d\to\infty$. The observations have encoding lengths polynomial in $n$ and $d$.
\end{theorem}

\begin{proof}
\textbf{Construction.} Choose an even integer $n=\Theta_C(d^2/\log d)$, where the constant will be specified below. Consider the configuration $X^0=(x_i^0)_{i\in[n]}$ of points on the moment curve,
\[
 x_i^0=(i,i^2,\ldots,i^d)\in\R^d.
\]
Every $d+1$ of these points are affinely independent by the Vandermonde determinant. Let $M$ be a perfect matching on $[n]$, and let $A^0$ be the squared distance matrix of $X^0$. For each matching edge $\{i,j\}\in M$, introduce a nonzero rational perturbation $\xi_{ij}$ and set
\[
 A_{ij}=\begin{cases}
 A^0_{ij}+\xi_{ij}, & \{i,j\}\in M,\\
 A^0_{ij}, & \text{otherwise}.
 \end{cases}
\]
Since only the matching observations are changed, $\err_A(X^0)=|M|=n/2$. In particular, $\OPT_d(A)\le n/2$. We choose the perturbations to satisfy two properties:
\begin{enumerate}
 \item For every subset of at most $d+1$ points, the observed squared distances admit an affinely independent realization in $\R^d$.
 \item For every subset of $d+2$ points containing a matching edge, the observed squared distances are not realizable in $\R^d$.
\end{enumerate}

\paragraph{Nonrealizability of larger subsets.}
Fix a set $S$ of $d+2$ points containing at least one matching edge. Let
\[
 C_S^0=\begin{pmatrix}0&\mathbf1^\top\\\mathbf1&A^0[S,S]\end{pmatrix}
\]
be its Cayley--Menger matrix before perturbation. We show that $C_S^0$ has nullity one. Let $(c,v)$ be a vector in the kernel, with $v=(v_i)_{i\in S}$. Then
\[
 \sum_{i\in S}v_i=0,\qquad c\mathbf1+A^0[S,S]v=0.
\]
For each $j\in S$, expanding the squared distances gives
\[
 (A^0[S,S]v)_j
 =\sum_{i\in S}v_i\|x_i^0\|^2
 -2\left\langle x_j^0,\sum_{i\in S}v_i x_i^0\right\rangle.
\]
Since this expression is independent of $j$, subtracting the expressions for any two indices $j,h\in S$ gives
\[
 \left\langle x_j^0-x_h^0,\sum_{i\in S}v_i x_i^0\right\rangle=0.
\]
The points indexed by $S$ affinely span $\R^d$, so $\sum_{i\in S}v_i x_i^0=0$. Conversely, every affine dependence among these points determines a kernel vector with $c=-\sum_{i\in S}v_i\|x_i^0\|^2$. Since $S$ has $d+2$ points and every $d+1$ of them are affinely independent, their space of affine dependencies has dimension one. Hence $C_S^0$ has nullity one. Furthermore, every coordinate $v_i$ of a nonzero dependence is nonzero; otherwise at most $d+1$ points would be affinely dependent.

Let $w=(c,v)$ span the kernel of $C_S^0$. Since $C_S^0$ is symmetric and has nullity one, its adjugate satisfies
\[
 \operatorname{adj}(C_S^0)=\mu ww^\top
\]
for some nonzero scalar $\mu$. Indeed, the adjugate is nonzero because the matrix has rank one less than its size, and each column and row lies in its one-dimensional kernel. Now choose a matching edge $\{i,j\}\subseteq S$. Let $C_S(\boldsymbol{\xi})$ be the Cayley--Menger matrix after perturbation, where $\boldsymbol{\xi}$ denotes the vector of all matching perturbations. Differentiating its determinant with respect to the symmetric entry corresponding to $\{i,j\}$ gives
\[
 \left.\frac{\partial\det C_S(\boldsymbol{\xi})}{\partial\xi_{ij}}\right|_{\boldsymbol{\xi}=0}
 =2\mu v_i v_j\ne0.
\]
Thus $\det C_S(\boldsymbol{\xi})$ is a nonzero polynomial in the matching perturbations. This holds for every $S$ containing a matching edge. If we choose perturbations that avoid the zero sets of all these polynomials, every such Cayley--Menger matrix has nonzero determinant. The Cayley--Menger determinant of $d+2$ points in $\R^d$ is zero, so the second property follows.

\paragraph{Choosing small rational perturbations.}
We now show that both properties can be achieved with rational perturbations of polynomial bit length. Set
\[
 U=\max\{1,\max_{i,j\in[n]}\|x_i^0-x_j^0\|^2\}\le dn^{2d}.
\]
By affine independence, every anchored Gram matrix $G$ of a subset of between two and $d+1$ original points is positive definite. Its entries are integers of absolute value at most $U$, so its determinant is a positive integer and hence at least one. Every eigenvalue is at most $dU$, and therefore
\[
 \lambda_{\min}(G)\ge(dU)^{-d}.
\]
If each squared distance changes by at most $\zeta$, each entry of the Gram matrix changes by at most $3\zeta/2$, by \eqref{eq:gram}. The spectral norm of this perturbation is at most $3d\zeta/2$. Set
\[
 \zeta=\frac{1}{4d(dU)^d}.
\]
Since $3d\zeta/2<(dU)^{-d}$, every perturbation in the interval $(0,\zeta)$ preserves positive definiteness of all these Gram matrices. The bit length of $\zeta$ is polynomial in $d$ and $\log n$.

For the second property, let $P(\boldsymbol{\xi})$ be the product of the polynomials $\det C_S(\boldsymbol{\xi})$ over all sets $S$ of size $d+2$ containing a matching edge. This product is nonzero and has total degree at most
\[
 D_0=(d+3)\binom{n}{d+2}.
\]
Choose $Q=D_0+1$ and consider the rational grid
\[
 \mathcal G=\left\{\frac{j\zeta}{Q+1}:j\in[Q]\right\}.
\]
A nonzero polynomial of total degree at most $D_0$ cannot vanish on the entire Cartesian product $\mathcal G^{|M|}$ when $Q>D_0$. This follows by induction on the number of variables using the univariate root bound. Hence some choice of perturbations from $\mathcal G^{|M|}$ satisfies $P(\boldsymbol{\xi})\ne0$. All of them are positive and smaller than $\zeta$, so both properties hold. Since $\log Q=O(d\log n)$, the perturbations and resulting observations have polynomial bit length. Singleton subsets are trivially realizable.

\paragraph{Cost of a reconstruction.}
We next analyze \cref{alg:pivots} on the constructed instance. For sufficiently large $d$, we have $n>d+2$ and $D=d$. Suppose the algorithm has selected $s$ anchors, including the initial anchor, with $1\le s\le d$. By the first property, the observed distances of these anchors together with any other point admit an affinely independent realization. Inductively, the selected anchors have been realized with all their mutual observations satisfied. For each point outside the anchor set, the update formulas recover its projection onto the anchor span and its squared residual norm in the corresponding realization. This residual norm is strictly positive by affine independence. Since the $s$ anchors span $s-1<d$ dimensions, no final-coordinate freezing rule is applied during these steps. Thus every point outside the anchor set remains active and all its distances to the anchors agree with the observations.

Consequently, the first $d$ anchors, including the initial anchor, are selected uniformly without replacement. Let $S$ be their index set. By the geometric invariant of \cref{sec:pivot-algorithm}, their observed distances to all points remain satisfied during subsequent updates. Suppose $S$ contains a matching edge. For any two distinct points $i,j\notin S$, if their mutual observation were also satisfied, every observed squared distance on $S\cup\{i,j\}$ would be realized in $\R^d$. This contradicts the second property, since $S\cup\{i,j\}$ has $d+2$ points and contains a matching edge. Therefore, every pair outside $S$ is a disagreement in the output, and the cost is at least $\binom{n-d}{2}$. The optimum is positive by the second property and at most $n/2$ by construction. The approximation ratio is thus at least
\[
 \frac{\binom{n-d}{2}}{n/2}=\Omega(n),
\]
using $n/d\to\infty$.

\paragraph{Independent runs.}
Condition on the first $j$ anchors containing no matching edge. Exactly $j$ of the remaining $n-j$ points are matched to these anchors, so the probability that the next anchor does not complete a matching edge is $(n-2j)/(n-j)$. The probability that a uniform set of $d$ anchors contains no matching edge is
\begin{align*}
 \prod_{j=0}^{d-1}\frac{n-2j}{n-j}
 &=\prod_{j=0}^{d-1}\left(1-\frac{j}{n-j}\right)\\
 &\le\exp\left(-\sum_{j=0}^{d-1}\frac{j}{n-j}\right)\\
 &\le\exp\left(-\frac{d(d-1)}{2n}\right).
\end{align*}
Choose the even integer $n$ so that
\[
 n=(c_C+o(1))\frac{d^2}{\log d},\qquad
 c_C=\frac{1}{8(C+1)}.
\]
Then $n^C\exp(-d(d-1)/(2n))=o(1)$, since $\log n=(2+o(1))\log d$ and $d(d-1)/(2n)=(4(C+1)+o(1))\log d$. By the union bound, with probability $1-o(1)$, every run among at most $n^C$ independent runs selects an anchor set containing a matching edge. Even the best output then has approximation ratio $\Omega(n)=\Omega_C(d^2/\log d)$.
\end{proof}

At $n=\Theta_C(d^2/\log d)$, the upper bound in \cref{thm:dim} is $O(d^2/\log d)$. Thus the lower bound matches it on this family of instances.